\documentclass{svjour2}  

\usepackage{bbm}
\usepackage{amsmath}
\usepackage{graphicx}
\usepackage{epsfig}

\usepackage{amsfonts}
\usepackage{amssymb}
\usepackage{bbold} 
\usepackage{amsbsy}

\def\e{\kern+.6ex\lower.42ex\hbox{$\scriptstyle \iota$}\kern-1.20ex e}

\journalname{Few-Body Syst}

\begin{document}

\title{A Three-Dimensional Treatment of the Three-Nucleon Bound 
State\thanks{Dedicated to Prof. H. Wita{\l}a on the occasion of his
  60th birthday}}
\author{J. Golak, K. Topolnicki, R. Skibi\'nski, 
  W. Gl\"ockle, H. Kamada, and A.~Nogga} 

\institute{
J.~Golak \and K.~Topolnicki \and R.~Skibi\'nski 
\at M. Smoluchowski Institute of Physics, Jagiellonian University, PL-30059 Krak\'ow, Poland 
\and W.~Gl\"{o}ckle
\at Institut f\"ur Theoretische Physik II, Ruhr-Universit\"at Bochum, D-44780 Bochum, Germany
\and H.~Kamada
\at Department of Physics, Faculty of Engineering, Kyushu Institute of Technology, 1-1 Sensuicho Tobata, Kitakyushu 804-8550, Japan
\and A.~Nogga 
\at Forschungszentrum J\"ulich, Institut f\"ur Kernphysik, Institute for Advanced Simulation 
and J\"ulich Center for Hadron Physics, D-52425 J\"ulich, Germany
}


\maketitle

\begin{abstract}
  Recently a formalism for a direct treatment of the Faddeev equation 
  for the three-nucleon (3N) bound state in three dimensions has been proposed.
  It relies on an operator representation of the Faddeev component 
  in the momentum space
  and leads to a finite set of coupled equations for
  scalar functions which depend only on three variables. 
  In this paper we provide further elements of this formalism and show 
  the first numerical results for chiral NNLO nuclear forces.
\end{abstract}



\section{Introduction}

Chiral three-nucleon (3N) forces at next-to-next-to-next-to-leading order (N3LO)
(see for example Refs.~\cite{N3LO,N3LO2}) 
have a very rich spin and momentum structure, so calculations of the 3N system based
on standard partial wave representations are very tedious~\cite{tritonn3lo}.
Recently, the Faddeev equations for the 3N system have been formulated 
directly with vector variables \cite{Bayegan:2007ih,Bayegan:2011,Bayegan:2012,2N3N}.
In contrast to Refs.~\cite{Bayegan:2007ih,Bayegan:2011,Bayegan:2012}, where
spin and isospin couplings are treated explicitly,
the key ingredient in Ref.~\cite{2N3N} 
is an operator representation of the 3N bound state. 
Such an operator form was given long time ago in Ref.~\cite{gerjuoy} and later 
re-derived in Ref.~\cite{fach04}.

Following this formulation, one represents two-nucleon (2N) t-operators,
 3N forces and Faddeev components of the 3N wave function as linear combinations involving products of
scalar functions and scalar operators built from the spin operators and
momentum vectors. By simple manipulations it is later possible
to remove the spin operators, which leads to relations among scalar functions
of momentum vectors only. In this paper we provide further elements of this 
formalism.

An alternative formulation of the 3N bound state equation
is discussed in Sect.~2. There we introduce an additional set of scalar coefficients 
required by the modified integral kernel.

In Sect.~3 we consider the full 3N bound state and show how its scalar
coefficients can be obtained from the scalar functions used to represent 
the Faddeev component. In the same section we discuss the normalization 
of the full wave function, as well as the expectation values of the kinetic energy
and the 3N potential energy.

Details of our numerical performance and some numerical results are provided in 
Sect.~4, where we use a simpler and more efficient way of introducing 3N forces.
We end with a brief summary and outlook in Sect.~5.

\section{An alternative form of the Faddeev equation using the operator expansion}

For the convenience of the reader we first remind of the building blocks of our formalism.

Working with the isospin concept, nucleons are treated as identical particles
so one needs just one equation for the Faddeev component $| \psi \rangle$. 
In Ref.~\cite{2N3N} this equation is written in the following form
\begin{eqnarray}
| \psi \rangle = G_0 t P | \psi \rangle  + ( 1 + G_0 t ) G_0 V^{(1)} ( 1 + P)  |\psi \rangle ~,
\label{faddeev}
\end{eqnarray}
where $ G_0 \equiv \frac1{E-H_0}$ is the free 3N propagator for the internal 3N energy $E$,
$t$ is the 2N transition operator for nucleons in the $(2,3)$ subsystem
and $P = P_{12}P_{23} + P_{13}P_{23}$ is the permutation operator.
The term $V^{(1)}$ is that part of the 3N force, 
which is symmetrical under exchange of nucleons 2 and 3. 
Of course, the choice of the subsystem pair $(2,3)$ is arbitrary. 

Alternatively, the Faddeev equation can be written without the transition matrix 
element $t$ as
\begin{eqnarray}
| \psi \rangle = G_0 V ( 1 + P ) | \psi \rangle + G_0 V^{(1)} ( 1 + P) |\psi \rangle ~,
\label{faddeev2}
\end{eqnarray}
using directly the 2N potential $V$ for nucleons 2 and 3.

In order to account for charge independence and charge symmetry
breaking, matrix elements of the 2N potential between the eigenstates 
of the 2N isospin operator
are given as
\begin{eqnarray}
\langle t m_t \mid V \mid t' m_{t'} \rangle  = \delta_{tt'}  \, \delta_{ m_t m_{t'}  } \, V^{ t m_t} \, .
\label{eq:isov}
\end{eqnarray}
It is well known \cite{wolfenstein} that the most general 2N force 
$ V^{ t m_t} $ has in the momentum space the following form
\begin{eqnarray}
\langle \vec p \mid V^{ t m_t} \mid \vec p' \rangle 
= \sum_{j=1}^6 v_j^{ t m_t} ( \vec p,\vec p') \;  w_j(\vec \sigma_{(2)},
\vec \sigma_{(3)}, \vec p,\vec p'),
\label{vexpand}
\end{eqnarray}
with $v_j^{ t m_t} ( \vec p,\vec p')$ being scalar functions of the initial ($\vec p'$)
and final ($\vec p$) relative Jacobi momenta. 
We choose the same spin-momentum scalar operators $w_j (\vec \sigma_{(2)},
\vec \sigma_{(3)}, \vec p,\vec p')$ as in Ref.~\cite{2n3d}
\begin{eqnarray}
w_1 (\vec \sigma_{(2)}, \vec \sigma_{(3)}, \vec p,\vec p') & = &  \mathbb{1} \, , \nonumber \\
w_2 (\vec \sigma_{(2)}, \vec \sigma_{(3)}, \vec p,\vec p') & = &  \vec \sigma_{(2)} \cdot \vec \sigma_{(3)} \, , \nonumber \\
w_3 (\vec \sigma_{(2)}, \vec \sigma_{(3)}, \vec p,\vec p') & = &  -i ( \vec \sigma_{(2)} + \vec \sigma_{(3)} \, ) \cdot ( \vec p \times \vec p')  \, , \nonumber \\
w_4 (\vec \sigma_{(2)}, \vec \sigma_{(3)}, \vec p,\vec p') & = &  \vec \sigma_{(2)} \cdot ( \vec p \times \vec p') \, \vec \sigma_{(3)} \cdot ( \vec p \times \vec p') \, , \nonumber \\
w_5 (\vec \sigma_{(2)}, \vec \sigma_{(3)}, \vec p,\vec p') & = &  \vec \sigma_{(2)} \cdot ( \vec p + \vec p') \, \vec \sigma_{(3)} \cdot ( \vec p + \vec p') \, , \nonumber \\
w_6 (\vec \sigma_{(2)}, \vec \sigma_{(3)}, \vec p,\vec p') & = &  \vec \sigma_{(2)} \cdot ( \vec p - \vec p') \, \vec \sigma_{(3)} \cdot ( \vec p - \vec p') \, .
\label{eq:3}
\end{eqnarray}
(Note this set is different from the one used in \cite{2N3N}.)
Also for the transition operator $t^{ t m_t}$, generated by the Lippmann-Schwinger equation,
\begin{eqnarray}
 t^{ t m_t} = V^{ t m_t} + V^{ t m_t} G_0^{2N} t^{ t m_t} \, ,
\label{eq:lse}
\end{eqnarray}
the same expansion is valid but the corresponding scalar functions 
$t_j^{ t m_t} ( \vec p,\vec p')$ depend also on the 2N energy, $E_{2N}$. 

In the next step, $| \psi \rangle $, $V$, $t$ and $V^{(1)}$ are expanded into the three possible 3N isospin states,
$ \mid \left( t \frac12   \right) T \rangle $, 
given in terms of the subsystem isospin $t$ and the total 3N isospin $T$:
\begin{eqnarray}
| \psi \rangle &=& \sum_{tT} \Bigg| \left( t\frac{1}{2}\right) T\Bigg\rangle \, | \psi_{tT} \rangle \, , \\
V &=& \sum_{t T, t' T' } \Bigg| \left( t\frac{1}{2}\right) T \Bigg\rangle \, V_{t T, t' T' } \, 
\Bigg\langle \left( t'\frac{1}{2}\right) T' \Bigg| \, , \\
t &=& \sum_{t T, t' T' } \Bigg| \left( t\frac{1}{2}\right) T\Bigg\rangle \, t_{t T, t' T' } \, \Bigg\langle \left( t'\frac{1}{2}\right) T' \Bigg| \, , \\
V^{(1)} &=& \sum_{t T, t' T' } \Bigg| \left( t\frac{1}{2}\right) T\Bigg\rangle \, V^{(1)}_{t T, t' T' } \, \Bigg\langle \left( t'\frac{1}{2}\right) T' \Bigg| \, .
\end{eqnarray}
Consistent with (\ref{eq:isov}), we assume that
\begin{eqnarray}
V_{t T, t' T' } &=& \delta_{t t'} \, V_{t T T'} \, ,
\label{isov2}
\end{eqnarray}
\begin{eqnarray}
t_{t T, t' T' } &=& \delta_{t t'} \, t_{t T T'} \, .
\end{eqnarray}
We consider only such 3N forces which conserve the total
3N isospin:
\begin{eqnarray}
V^{(1)}_{t T, t' T' } &=& \delta_{T T'} V^{(1)}_{t t' T}\, .
\label{V3N}
\end{eqnarray}
The isospin matrix elements of the permutation operator $P$ 
read~\cite{Witala:2009ws1}
\begin{eqnarray}
\langle \left( t \frac{1}{2}\right) T \mid P \mid \left( t' \frac{1}{2}\right) T' \rangle
= \delta_{TT'}
 F_{ tt' T} ( P_{12}^{sm} P_{23}^{sm}
 + ( -)^{t+t'} P_{13}^{sm} P_{23}^{sm}) ~,
\label{PME}
\end{eqnarray}
where $F_{ tt' T}$ are geometrical factors and $ P_{ij}^{sm}$
acts only in the spin and momentum spaces.
Using the above information, the Faddeev equation (\ref{faddeev2}) 
becomes 
\begin{eqnarray}
| \psi_{tT} \rangle = G_0 \sum_{T'} V_{ tTT'} \, \sum_{t'} \left( \delta_{t t'} \, + \, 
      F_{t t' T'} \left( P_{12}^{sm} P_{23}^{sm} + ( -)^{t+t'} P_{13}^{sm} P_{23}^{sm}\right) 
                   \right) \, | \psi_{t'T'} \rangle \nonumber \\
          + \ G_0 \sum_{t'} V^{(1)}_{ t t' T } \, \sum_{t''} 
     \left( \delta_{t' t''} \, + \,   F_{t' t'' T} \left( P_{12}^{sm} P_{23}^{sm} 
  + ( -)^{t'+t''} P_{13}^{sm} P_{23}^{sm}\right) 
        \right) \, | \psi_{t''T} \rangle \, .
\label{eq:272}
\end{eqnarray}
(The isospin projected equation (\ref{faddeev}) can be found in Ref.~\cite{2N3N}.)
As expected, the charge independence and charge symmetry breaking in the 2N
force leads to the coupling of total isospin $T=\frac12$ and $\frac32$.

In the following we will consider the above equation in the momentum space 
spanned by the two relative Jacobi momenta $ \vec p $ and $ \vec q$. 
In Eq.~(\ref{faddeev}) we encounter the off-shell 2N $t$-matrix
\begin{eqnarray}
\langle \vec p \vec q| t_{t TT'}| \vec p' \vec q'\rangle = \delta( \vec q -
\vec q') t_{tTT'} ( \vec p, \vec p',E_q)
\end{eqnarray}
with $E_q \equiv E_{2N} \equiv E - \frac {3} {4m} q^2$.
In the case of Eq.~(\ref{faddeev2}) we 
only need the 2N potential matrix elements
\begin{eqnarray}
\langle \vec p \vec q| V_{t TT'}| \vec p' \vec q'\rangle = \delta( \vec q -
\vec q') V_{tTT'} ( \vec p, \vec p') \, .
\end{eqnarray}
The matrices for the permutation operators read~\cite{Gl83}
\begin{eqnarray}
\langle \vec p \vec q| P_{12}^{sm} P_{23}^{sm}| \vec p' \vec q'\rangle & = &
\delta( \vec p - \vec \pi( \vec q, \vec q')) \delta( \vec p' -
\vec \pi'( \vec q, \vec q'))P_{12}^{s} P_{23}^{s}
\label{PME2}
\end{eqnarray}
\begin{eqnarray}
\langle \vec p \vec q| P_{13}^{sm} P_{23}^{sm}| \vec p' \vec q'\rangle &=&
\delta( \vec p + \vec \pi( \vec q, \vec q')) \delta( \vec p' +\vec
\pi'( \vec q, \vec q'))P_{13}^{s} P_{23}^{s}
\label{PME22}
\end{eqnarray}
with
\begin{eqnarray}
\vec \pi( \vec q, \vec q') = \frac{1}{2} \vec q + \vec q'\\
\vec \pi'( \vec q, \vec q')= -\vec q - \frac{1}{2} \vec q' ~.
\label{PME3}
\end{eqnarray}

As pointed out in Ref.~\cite{2N3N}, it is crucial for the present formulation
to write the Faddeev amplitude $ \psi_{tT}( \vec p, \vec q) \equiv 
\langle \vec p \vec q \mid \psi_{tT} \rangle $ in the operator form~\cite{fach04}.
It reads
\begin{eqnarray}
\psi_{tT}( \vec p, \vec q) = \sum_{ i = 1}^ 8 \phi_{ tT}^ {(i)} (
\vec p,\vec q) \; O_i \vert \chi^ m \rangle \, ,
\label{eq:38}
\end{eqnarray}
where $|\chi^ m \rangle = | ( 0 \frac{1}{2}) \frac{1}{2} m \rangle $ is a
specific state in which the three spin-$1/2$ states are coupled to the 
total angular momentum
quantum numbers of the 3N bound state. The functions  $\phi_{ tT}^ {(i)} (
\vec p,\vec q) \equiv \phi_{ tT}^ {(i)} (
|\vec p|,|\vec q|,\hat{\vec p} \cdot \hat{\vec q} )$ are scalar functions and the operators $ O_i$ are
given as
\begin{eqnarray}
O_1 & = &  1 \, ,\nonumber \\
O_2 & = &  \frac{1}{ \sqrt{3}} \; \vec \sigma( 23) \cdot \vec \sigma_{(1)} \, ,\nonumber \\
O_3 & = &  \sqrt{\frac{3}{2}} \frac{1}{i} \; \vec \sigma_{(1)} \cdot \left( {\bf \hat p} \times {\bf \hat q} \, \right)\, , \nonumber \\
O_4 & = &  \frac{1}{ \sqrt{2}} \Big[ i \vec \sigma( 23)  \cdot \left( {\bf \hat p} \times {\bf \hat q} \, \right)
 - \left( \vec \sigma_{(1)} \times \vec \sigma( 23) \, \right) \cdot ({\bf \hat p} \times {\bf \hat q} \, ) \Big] \, , \nonumber \\
O_5 & = &  \frac{1}{i} \Big[ \vec \sigma( 23)  \cdot ({\bf \hat p} \times {\bf \hat q} \, ) \ - \ \frac{ i}{2} \left( \vec \sigma_{(1)} \times
\vec \sigma( 23) \right)   \cdot ({\bf \hat p} \times {\bf \hat q} \, ) \, \Big]\, ,  \nonumber \\
O_6 & = &  \sqrt{\frac{3}{2}} \Big( \vec \sigma( 23) \cdot {\bf \hat p} \; \vec \sigma_{(1)} \cdot {\bf \hat p}
 - \frac{1}{3} \, \vec \sigma( 23) \cdot \vec \sigma_{(1)}  \Big)\, , \nonumber \\
O_7 & = &   \sqrt{\frac{3}{2}}  \Big(\vec \sigma( 23) \cdot {\bf \hat q} \;
\vec \sigma_{(1)} \cdot {\bf \hat q} - \frac{1}{3} \, \vec \sigma( 23) \cdot \vec \sigma_{(1)}  \Big)\, , \nonumber \\
O_8 & = &  \frac{3}{2} \frac{ 1}{ \sqrt{5}}  \Big( \vec \sigma( 23)
\cdot {\bf \hat q} \; \vec \sigma_{(1)} \cdot {\bf \hat p} + \vec \sigma( 23) \cdot
{\bf \hat p} \vec \sigma_{(1)} \cdot {\bf \hat q}\cr &  - &  \frac{2}{3} {\bf \hat p}
\cdot {\bf \hat q} \vec \sigma( 23) \cdot \vec \sigma_{(1)} \Big) ~,
\label{o18}
\end{eqnarray}
where $ \vec \sigma(23) \equiv \frac{1}{2} ( \vec \sigma_{(2)} - \vec \sigma_{(3)})$.
We use unit vectors, $\bf \hat p$ and $\bf \hat q$.
Note that we actually use the second set of operators from Ref.~\cite{2N3N}.
Note also that several misprints 
in Eq.~(31) of Ref.~\cite{2N3N} has been now corrected using \cite{fach04}.

As already mentioned, we can use the most general operator structure
of the 2N t-matrix $ t_{tTT'}(\vec p,\vec p',E_q)$, which is given by
\begin{eqnarray}
t_{ tTT'}( \vec p, \vec p',E_q) = \sum_{ j=1}^ 6 t_{ t T T'}^ {
(j)} ( \vec p, \vec p',E_q) \; w_j (\vec \sigma_{(2)}, \vec
\sigma_{(3)}, 
\vec p, \vec p').
\label{eq:55}
\end{eqnarray}
Here the matrix elements  $t_{ t T T'}^ { (j)} ( \vec p, \vec p',E_q) $ are scalar
functions and the operators $w_j( \vec \sigma_{(2)}, \vec \sigma_{(3)}, \vec p,
\vec p')$ are  given in Eq.~(\ref{eq:3}).

\noindent
In principle, the 3N force operator $V_{ tt'T}^{(1)} ( \vec p, \vec q, \vec
p', \vec q')$ also allows for an expansion
\begin{eqnarray}
V_{ tt'T}^{(1)} ( \vec p, \vec q, \vec p', \vec q') = \sum_l v_{ t
t' T}^{(l)} (\vec p, \vec q, \vec p', \vec q') \; \Omega_l(\vec
\sigma_{(1)}, \vec \sigma_{(2)}, \vec \sigma_{(3)}, \vec p, \vec q, \vec p', \vec q')
\label{eq:62}
\end{eqnarray}
with a strictly finite number of terms~\cite{hermann} but for computational reasons we 
treat it as one object and do not use any expansion. This and the fact that our integral 
kernel is simpler than in Ref.~\cite{2N3N}, leads to a substantial 
simplification and a reduction of the number of scalar coefficients 
introduced in Ref.~\cite{2N3N}.

Using exactly the same algebra as in Ref.~\cite{2N3N}, 
we arrive at the final form of the Faddeev equation (\ref{faddeev2}), which corresponds 
to Eq.~(50) in Ref.~\cite{2N3N}:
\begin{eqnarray}
\lefteqn{\sum_{ i = 1}^ 8 C_{ki}(\vec p \vec q,\vec p \vec q )  \; 
\phi_{ tT}^ {(i)} ( \vec
p,\vec q)  =} \cr
& & \frac{1}{ E-\frac{p^2}{m} - \frac{3}{4m} q^2} \; \int
d^3 q' \; \sum_{T'} \sum_{ k' = 1}^ 8 \cr 
& & 
\sum_{ j=1}^ 6 \Big( 
v_{t T T'}^ { (j)} ( \vec p, \vec \pi ( \vec q,\vec q')) D_{kjk'}
(\vec p \vec q, \vec p \vec \pi( \vec q,\vec q'), \vec \pi'( \vec
q,\vec q') \vec q') \cr
&+&  (-)^t v_{ t T T'}^{ (j)} ( \vec p, - \vec \pi ( \vec
q,\vec q') )    D_{kjk'}^{'} (\vec p \vec q, \vec p (-\vec
\pi( \vec q,\vec q')), \vec \pi'( \vec q,\vec q') \vec q')\Big) \cr 
& & \times  \sum_{t'}F_{ tt' T'}\phi_{ t'T'}^ {(k')} ( \pi'( \vec q,\vec
q') \vec q') \cr
&+& \frac{1}{ E-\frac{p^2}{m} - \frac{3}{4m} q^2}
\int d^3 p' \sum_{T'} \sum_{ j = 1}^6 \sum_{ k' = 1}^8
v_{ t T T'}^{ (j)}  (\vec p, \vec p') 
 L_{k j k' } ( \vec p \vec q, \vec p \vec p' , \vec p' \vec q )\phi_{ tT'}^ {(k')} 
( \vec p',\vec q )\cr
&+& \frac{1}{ E-\frac{p^2}{m} - \frac{3}{4m} q^2}
\int d^3 p' d^3 q' \sum_{t'} 
\sum_{ k' = 1}^ 8 E_{kk'}^{ t t' T} ( \vec p \vec
q, \vec p \vec q \vec p' \vec q', \vec p' \vec q')\phi_{ t'T}^ {(k')} 
( \vec p',\vec q')\cr
&+& \frac{1}{ E-\frac{p^2}{m} -
\frac{3}{4m} q^2} \int d^3 q' d^3 q'' \sum_{t'} \,
\sum_{ k' = 1}^ 8 
\Big( 
G_{kk'}^{ t t' T} ( \vec p \vec q, \vec p \vec q \vec
\pi(\vec q', \vec q'') \vec q', \vec \pi'(\vec q', \vec q'')
\vec q'')  \cr
 & + &   (-)^ {t'}
G_{kk'}^{' \, t t' T} ( \vec p \vec q, \vec p
\vec q (-\vec \pi(\vec q', \vec q'')) \vec q', \vec \pi'(\vec
q', \vec q'') \vec q'') \Big) \cr & &
\times \sum_{t''} F_{ t't'' T} \phi_{
t''T}^ {(k')} ( \vec \pi'( \vec q',\vec q''), \vec q'')\, .
\label{eq:790}
\end{eqnarray}
We use here seven sets of scalar coefficients: 
$ C_{ki} $,
$ L_{kji}$,
$ D_{kji}$,
$ D_{kji}^{'}$,
$ E_{k i}^{t t' T}$,
$ G_{k i}^{t t' T}$
and $G_{k k'}^{' \, t t' T}$.
$ C_{ki} $,
$ D_{kjk'}$ and
$ D_{kjk'}^{'}$
are exactly the same as in Ref.~\cite{2N3N},
where they appear in the $ G_0 t P \mid \psi \rangle $ 
part of the Faddeev kernel. 
The coefficients 
$E_{k k'}^{t t' T}$,
$G_{k k'}^{t t' T}$ and
$G_{k k'}^{' \, t t' T}$,
which are required to deal with the 3N force in the Faddeev 
equation,
are slightly modified versions of the quantities appearing 
in Ref.~\cite{2N3N}, because we do not expand 
$V_{ tt'T}^{(1)} ( \vec p, \vec q, \vec p', \vec q') 
\equiv V_{ tt'T}^{(1)} (\vec \sigma_{(1)}, \vec \sigma_{(2)}, \vec \sigma_{(3)}, \vec p, \vec q, \vec p', \vec q') $:
\begin{eqnarray}
\lefteqn{E_{k k'}^{ tt'T} (\vec p \vec q , \vec p \vec q  \vec p' \vec q',
\vec p' \vec q') = } \cr
& & \sum_{m} \langle \chi^
m| O_k ( \vec \sigma_{(1)}, \vec \sigma_{(2)}, \vec \sigma_{(3)}, \vec
p,
\vec q) \nonumber \\
& & \times V_{ tt'T}^{(1)} (\vec \sigma_{(1)}, \vec \sigma_{(2)}, \vec \sigma_{(3)}, \vec p, \vec q, \vec p', \vec q')
\;  O_{k'} ( \vec
\sigma_{(1)}, \vec \sigma_{(2)}, \vec \sigma_{(3)}, \vec p', \vec
q')|\chi^ m
\rangle \, ,
\label{eq45}
\end{eqnarray}
\begin{eqnarray}
\lefteqn{G_{k k'}^{ tt'T}   (\vec p \vec q ,\vec p \vec q \vec p' \vec q', \vec p''
\vec q'') = } \cr
& & -\frac{1}{2}\sum_{m}  \langle \chi^ m|  O_k ( \vec
\sigma_{(1)}, \vec \sigma_{(2)}, \vec \sigma_{(3)}, \vec p, \vec q)
V_{ tt'T}^{(1)} (\vec \sigma_{(1)}, \vec \sigma_{(2)}, \vec \sigma_{(3)}, \vec p, \vec q, \vec p', \vec q')\cr 
& &
\times O_{k'} ( \vec \sigma_{(2)}, \vec \sigma_{(3)}, \vec \sigma_{(1)}, \vec
p'', \vec q'') \;
  ( 1 + \vec \sigma(2,3) \cdot \vec \sigma_{(1)})| \chi^ m \rangle  \, ,
\label{eq47}
\end{eqnarray}
\begin{eqnarray}
\lefteqn{G_{k k'}^{' \, t t' T}  (\vec p \vec q ,\vec p \vec q \vec p' \vec q', \vec
p'' \vec q'') = }\cr
&  & \frac{1}{2} \sum_{m}  \langle \chi^ m|  O_k ( \vec
\sigma_{(1)}, \vec \sigma_{(2)}, \vec \sigma_{(3)}, \vec p, \vec q)
V_{ tt'T}^{(1)} (\vec \sigma_{(1)}, \vec \sigma_{(2)}, \vec \sigma_{(3)}, \vec p,
\vec q, \vec p', \vec q')\cr 
& &  \times O_{k'} ( \vec \sigma_{(3)}, \vec
\sigma_{(2)} \vec \sigma_{(1)} \vec p'', \vec q'')\; 
  ( 1 - \vec \sigma(2,3) \cdot \vec \sigma_{(1)})| \chi^ m \rangle  \, .
\label{eq48}
\end{eqnarray}
For the relatively simple chiral NNLO 3N force \cite{epel02},
which we use in this paper, the coefficients $E_{k k'}^{ tt'T}$ , $G_{k k'}^{ tt'T}$ and $G_{k k'}^{' \, t t' T}$  from Eq.~(\ref{eq45}) , (\ref{eq47}) and (\ref{eq48})
can be easily calculated for the whole 3N force.


The only new coefficients, $ L_{kjk'}$, are due to the 
$ G_0 V \, | \psi \rangle $ term in Eq.~(\ref{faddeev2}):
\begin{eqnarray}
\lefteqn{L_{kjk'} (\vec p \vec q,  \vec p' 
\vec p'',  \vec p''' \vec q''') = }\cr
& & \sum_{m} \langle \chi^ m|  O_k ( \vec \sigma_{(1)}, \vec
\sigma_{(2)}, \vec \sigma_{(3)}, \vec p, \vec q)  w_j(\vec \sigma_{(2)},
\vec \sigma_{(3)}, \vec p', \vec p'')\cr
 & & O_{k'} ( \vec \sigma_{(1)}, \vec
\sigma_{(2)}, \vec \sigma_{(3)}, \vec p''', \vec q''') 
| \chi^ m \rangle \, .
\label{eq43}
\end{eqnarray}
It is possible to further simplify Eq.~(\ref{eq:790}) because 
of the scalar nature of the $\phi$ functions and coefficients 
that one encounters there. 
To this aim it is convenient to choose our coordinate system
in such a way that 
\begin{flalign}
& \vec q = ( 0 , 0, q ) \, , & \nonumber \\
& \vec p = ( p \sqrt{1 - x^2 } , 0 , p x ) \, , &
\label{pqvec}
\end{flalign}
with $ x ={\bf \hat p} \cdot {\bf \hat q} $.


The $C_{ki}$ coefficients on the left-hand side of Eq.~(\ref{eq:790})
form a matrix with the determinant
\begin{eqnarray}
det(C) = \frac{2916}{5} \, \left| {\bf \hat p} \times {\bf \hat q}  \right|^{12} \,
\, = \frac{2916}{5} \, \left( 1 - x^2 \, \right)^6 \, .
\label{detc}
\end{eqnarray}
It means that the $C$ matrix might in principle cause numerical problems.
However, the application of the inverse matrix $C^{-1}$ to the both sides 
of Eq.~(\ref{eq:790}) and combining the $C^{-1}$ matrix elements with the
$L$, $D$, $D'$, $E$, $G$ and $G'$ coefficients leads to a scheme, which
proves numerically safe. Obviously, the case $ {\bf \hat p} \times {\bf \hat q} =0 $
has to be avoided in the numerical realization.


The 3N bound state Faddeev equation is solved by iteration \cite{stadler}, 
using a Lanczos type algorithm. This requires a choice of initial amplitudes 
$\phi_{ tT}^ {(i)} ( \vec p,\vec q)$, which guarantees the Fermi
character of the 3N system.
Now the operators $ O_2, O_3, O_6 $ and $ O_7 $ are odd under exchange of 
particles 2 and 3, whereas  $ O_1, O_4, O_5 $ and $ O_8 $ are even.
Thus the scalar functions must possess the following symmetry properties
(see Ref.~\cite{2N3N} for the derivation):
\begin{eqnarray}
\phi_{ tT}^ {(i)} (- \vec p,\vec q) & = &  (-)^ {t+1} \phi_{ tT}^ {(i)} ( \vec p,\vec q) \, , \, \, i=1,4,5,8 \, ,\nonumber \\
\phi_{ tT}^ {(i)} (- \vec p,\vec q) & = &  (-)^ {t} \phi_{ tT}^ {(i)} ( \vec p,\vec q)   \, , \, \, i=2,3,6,7 \, .
\label{eq:94} 
\end{eqnarray}

Before we embark on the description of our numerical performance, we will show how to obtain the operator
form for the full 3N bound state. This actually leads to yet another numerical scheme 
in comparison to Ref.~\cite{2N3N}.

\section{The Full 3N bound state}  

The total 3N state $| \Psi \rangle $ is obtained from the Faddeev 
component as $ | \Psi \rangle  = (1 + P) \, | \psi \rangle $.
If the isospin projected Faddeev component  
$ \psi_{tT}( \vec p, \vec q) \equiv
\langle \vec p\vec q|\psi_{tT}\rangle \equiv
\langle\, \vec p \, \vec q \, | \langle \left( t \frac12 \right) T | \psi \rangle $
is expanded (see Eq.~(\ref{eq:38})) 
using the expansion functions  $ \phi_{ tT}^ {(i)} (\vec p,\vec q) $, our task now is to find 
the expansion functions for 
$ A_{tT}( \vec p, \vec q) \equiv \langle \, \vec p \, \vec q \, | \langle \left( t \frac12 \right) T | P \psi \rangle $.
We use Eq.~(\ref{PME}) but rewrite
Eqs.~(\ref{PME2})--(\ref{PME3}) as 
\begin{eqnarray}
\langle\, \vec p \, \vec q \, | \, P_{12}^{sm} P_{23}^{sm}\, |\, {\vec p'} \, {\vec q'}\, \rangle & = &
\delta( {\vec p'} - \vec P_1 ( \vec p, \vec q ))
\delta( {\vec q'} - \vec Q_1 ( \vec p, \vec q )) P_{12}^{s} P_{23}^{s} \, , \nonumber \\
\langle \, \vec p \, \vec q \, |\, P_{13}^{sm} P_{23}^{sm}\, | \,{\vec p'} \,  {\vec q'}\, \rangle & = &
\delta( {\vec p'} - \vec P_2 ( \vec p, \vec q ))
\delta( {\vec q'} - \vec Q_2 ( \vec p, \vec q )) P_{13}^{s} P_{23}^{s}
\label{Pelements2}
\end{eqnarray}
with
\begin{eqnarray}
\vec P_1 ( \vec p, \vec q ) = -\frac{1}{2} \vec p - \frac34 \vec q \, , \\
\vec Q_1 ( \vec p, \vec q ) = \vec p - \frac{1}{2} \vec q \, ,  \\
\vec P_2 ( \vec p, \vec q ) = -\frac{1}{2} \vec p + \frac34 \vec q \, , \\
\vec Q_2 ( \vec p, \vec q ) = - \vec p - \frac{1}{2} \vec q \, .
\end{eqnarray}
Consequently,
we get
\begin{eqnarray}
A_{tT}( \vec p, \vec q) =  \int d^3 p'  \int d^3 q'  \; \sum_{t'}F_{ tt' T} \;
\left\{
\delta( {\vec p'} - \vec P_1 ( \vec p, \vec q ))
\delta( {\vec q'} - \vec Q_1 ( \vec p, \vec q )) P_{12}^{s} P_{23}^{s} \, \right. \nonumber \\
\left. + \,  (-)^{t+t'} \;
\delta( {\vec p'} - \vec P_2 ( \vec p, \vec q ))
\delta( {\vec q'} - \vec Q_2 ( \vec p, \vec q )) P_{13}^{s} P_{23}^{s}
\right\} \,
 \psi_{t'T}\left( {\vec p'} , {\vec q'} \, \right)  \nonumber \\
= \sum_{t'}F_{ tt' T} \;
\left\{  P_{12}^{s} P_{23}^{s} \, \psi_{t'T}\left( {\vec P}_1 , {\vec Q}_1  \, \right)
\, + \,
(-)^{t+t'} \; P_{13}^{s} P_{23}^{s} \, \psi_{t'T}\left( {\vec P}_2 , {\vec Q}_2  \, \right) \;
\right\}
\label{eq:332}
\end{eqnarray}
On the other hand we write $A_{tT}( \vec p, \vec q)$
directly in terms of some $\alpha_{ tT}^ {(i)} ( \vec p,\vec q)$ scalar coefficients
\begin{eqnarray}
A_{tT}( \vec p, \vec q) = \sum_{ i = 1}^ 8 \alpha_{ tT}^ {(i)} (
\vec p,\vec q) \; O_i \vert \chi^ m \rangle \, .
\label{eq:382}
\end{eqnarray}
Combining (\ref{eq:38}), (\ref{eq:332}) and (\ref{eq:382}) we get:
\begin{eqnarray}
\sum_{ i = 1}^ 8
\alpha_{ tT}^ {(i)} ( \vec p,\vec q) \;
O_i \left(
{\vec \sigma}_{(1)} ,
{\vec \sigma}_{(2)} ,
{\vec \sigma}_{(3)} ,
{\vec p} , {\vec q} \,
\right) \,
 \vert \chi^ m \rangle \, \nonumber \\
= \sum_{t'}F_{ tt' T} \;
\left\{  P_{12}^{s} P_{23}^{s} \,
\sum_{ i = 1}^ 8
\phi_{ t'T}^ {(i)} ( {\vec P}_1 , {\vec Q}_1 \, ) \;
O_i \left(
{\vec \sigma}_{(1)} ,
{\vec \sigma}_{(2)} ,
{\vec \sigma}_{(3)} ,
{\vec P}_1 , {\vec Q}_1  \,
\right) \,
 \vert \chi^ m \rangle \, \right. \nonumber \\
\left.
+ \,
(-)^{t+t'} \;
P_{13}^{s} P_{23}^{s} \,
\sum_{ i = 1}^ 8
\phi_{ t'T}^ {(i)} ( {\vec P}_2 , {\vec Q}_2 \, ) \;
O_i \left(
{\vec \sigma}_{(1)} ,
{\vec \sigma}_{(2)} ,
{\vec \sigma}_{(3)} ,
{\vec P}_2 , {\vec Q}_2  \,
\right) \,
 \vert \chi^ m \rangle \,
\right\}  \, .
\label{lll}
\end{eqnarray}

In order to get the $ \alpha_{ tT}^ {(i)} ( \vec p,\vec q) $ functions 
we first project Eq.~(\ref{lll}) from the
left with $ \langle \chi^ m|  O_k ( \vec \sigma_{(1)}, \vec \sigma_{(2)}, \vec
\sigma_{(3)}, \vec p, \vec q) $ and sum over $m$. We encounter again
the $C_{ki}$ coefficients 
together with the new coefficients
\begin{eqnarray}
W_{ki}\left( \vec p \vec q, {\vec P}_1  {\vec Q}_1 \, \right) = 
 \sum_{m}  \langle \chi^ m|  O_k ( \vec
\sigma_{(1)}, \vec \sigma_{(2)}, \vec \sigma_{(3)}, \vec p, \vec q) \nonumber \\
P_{12}^s P_{23}^s \,
 O_i ( \vec \sigma_{(1)}, \vec \sigma_{(2)}, \vec \sigma_{(3)}, {\vec P}_1 , {\vec Q}_1 \, )|\chi^ m \rangle \, ,
\label{eq422}
\end{eqnarray} 
\begin{eqnarray}
Z_{ki}\left( \vec p \vec q, {\vec P}_2  {\vec Q}_2 \, \right) = 
\sum_{m}  \langle \chi^ m|  O_k ( \vec
\sigma_{(1)}, \vec \sigma_{(2)}, \vec \sigma_{(3)}, \vec p, \vec q) \nonumber \\
P_{13}^s P_{23}^s \,
 O_i ( \vec \sigma_{(1)}, \vec \sigma_{(2)}, \vec \sigma_{(3)}, {\vec P}_2 , {\vec Q}_2 \, )|\chi^ m \rangle \, .
\label{eq423}
\end{eqnarray} 
In order to calculate 
$ W_{ki}\left( \vec p \vec q, {\vec P}_1  {\vec Q}_1 \, \right) $
and
$ Z_{ki}\left( \vec p \vec q, {\vec P}_2  {\vec Q}_2 \, \right) $
the following identities can be used, again following \cite{2N3N}
\begin{eqnarray}
\lefteqn{ P_{12}^{s} P_{23}^{s} \; O_k ( \vec \sigma_{(1)}, \vec \sigma_{(2)},
\vec \sigma_{(3)}, \vec p, \vec q ) | \chi^ m \rangle   = }\cr
& &  - \frac{1}{2}
O_k ( \vec \sigma_{(2)}, \vec \sigma_{(3)}, \vec \sigma_{(1)}, \vec p,\vec q)
( 1 + \vec \sigma(23)\cdot \vec \sigma_{(1)}) |\chi^ m \rangle \, .
\label{eq:68}
\end{eqnarray}
\begin{eqnarray}
\lefteqn{ P_{13}^{s} P_{23}^{s} \; O_k ( \vec \sigma_{(1)}, \vec \sigma_{(2)},
\vec \sigma_{(3)}, \vec p, \vec q ) | \chi^ m \rangle   = }\cr
& &  - \frac{1}{2}
O_k ( \vec \sigma_{(3)}, \vec \sigma_{(1)}, \vec \sigma_{(2)}, \vec p,\vec q)
( 1 - \vec \sigma(23)\cdot \vec \sigma_{(1)}) |\chi^ m \rangle \, .
\label{eq:69}
\end{eqnarray}
Then $W_{ki}$ and $Z_{ki}$ will be calculated as  
\begin{eqnarray}
W_{ki}\left( \vec p \vec q, {\vec P}_1  {\vec Q}_1 \, \right) = -\frac12 \,\sum_{m}  \langle \chi^ m|  O_k ( \vec
\sigma_{(1)}, \vec \sigma_{(2)}, \vec \sigma_{(3)}, \vec p, \vec q) \, \nonumber \\
\times O_i ( \vec \sigma_{(2)}, \vec \sigma_{(3)}, \vec \sigma_{(1)}, {\vec P}_1 , {\vec Q}_1 \, )
\, ( 1 + \vec \sigma(23)\cdot \vec \sigma_{(1)}) |\chi^ m \rangle \, ,
\label{eq422n}
\end{eqnarray} 
\begin{eqnarray}
Z_{ki}\left( \vec p \vec q, {\vec P}_2  {\vec Q}_2 \, \right) = -\frac12 \, \sum_{m}  \langle \chi^ m|  O_k ( \vec
\sigma_{(1)}, \vec \sigma_{(2)}, \vec \sigma_{(3)}, \vec p, \vec q) \, \nonumber \\
\times O_i ( \vec \sigma_{(3)}, \vec \sigma_{(1)}, \vec \sigma_{(2)}, {\vec P}_2 , {\vec Q}_2 \, )
\, ( 1 - \vec \sigma(23)\cdot \vec \sigma_{(1)}) |\chi^ m \rangle \, .
\label{eq423n}
\end{eqnarray}

Using the above definitions, Eq.~(\ref{lll}) can be written as
\begin{eqnarray}
\sum_{ i = 1}^ 8 
C_{ki}\left( \vec p \vec q,\vec p \vec q \, \right)  \,
\alpha_{ tT}^ {(i)} ( \vec p,\vec q) \; = \;
\sum_{t'} F_{ tt' T} \; \sum_{ l = 1}^8 
\left\{ 
W_{kl}\left( \vec p \vec q, {\vec P}_1  {\vec Q}_1 \, \right) \,
\phi_{ t'T}^ {(l)} ( {\vec P}_1 , {\vec Q}_1 \, ) \right. \nonumber \\
\left. + \;
(-)^{t+t'} \; Z_{kl}\left( \vec p \vec q, {\vec P}_2  {\vec Q}_2 \, \right) \,
\phi_{ t'T}^ {(l)} ( {\vec P}_2 , {\vec Q}_2 \, ) \,
\right\} \, .
\end{eqnarray}

The $ C_{ki}\left( \vec p \vec q,\vec p \vec q \, \right) $ are known analytically.
Thus we can easily find the inverse coefficients 
$ C^{-1}_{ki}\left( \vec p \vec q,\vec p \vec q \, \right) $, which obey
\begin{eqnarray}
\sum_{ k = 1}^ 8
C^{-1}_{mk}\left( \vec p \vec q,\vec p \vec q \, \right) \,
C_{ki}\left( \vec p \vec q,\vec p \vec q \, \right)  \, = \, \delta_{mi} \, .
\label{Cm1}
\end{eqnarray}
Using (\ref{Cm1}) we finally get
\begin{eqnarray}
\alpha_{ tT}^ {(m)} ( \vec p,\vec q) \; = \;
\sum_{t'} F_{ tt' T} \; 
\sum_{ k = 1}^8 
C^{-1}_{mk}\left( \vec p \vec q,\vec p \vec q \, \right)  \,
\sum_{ l = 1}^8 
\left\{ 
W_{kl}\left( \vec p \vec q, {\vec P}_1  {\vec Q}_1 \, \right) \,
\phi_{ t'T}^ {(l)} ( {\vec P}_1 , {\vec Q}_1 \, ) \right. \nonumber \\
\left. + \;
(-)^{t+t'} \; Z_{kl}\left( \vec p \vec q, {\vec P}_2  {\vec Q}_2 \, \right) \,
\phi_{ t'T}^ {(l)} ( {\vec P}_2 , {\vec Q}_2 \, ) \,
\right\} \, .
\label{alpha}
\end{eqnarray}
It is then clear that in order to calculate $\alpha_{ tT}^ {(m)} ( \vec p,\vec q)$
we need three-dimensional interpolations. In our numerical realization 
we use the cubic Hermite spline interpolation
method, described for example in Ref.~\cite{cubherm}.

Now we would like to pose the question of the 3N wave function normalization. 
The obvious starting point is
\begin{eqnarray}
1 = \langle \Psi \mid \Psi \rangle = 
\langle \psi  ( 1 + P ) \mid (1 + P ) \psi \rangle = 
3 \, \langle \psi  \mid \Psi \rangle \, .
\label{norm1}
\end{eqnarray}
The full 3N bound state $| \Psi \rangle $ can be written as
\begin{eqnarray}
| \Psi \rangle  = 
\sum_{t T} 
\int \! d^3 p 
\int \! d^3 q 
\mid  \vec p\, \vec q \, \rangle 
\mid \left( t \frac12 \right) T \rangle \nonumber \\
\sum_{ i = 1}^ 8
\beta_{t T}^{(i)} \left( {\vec p} , {\vec q} \, \right) \;
O_i \left(
{\vec \sigma}_{(1)} ,
{\vec \sigma}_{(2)} ,
{\vec \sigma}_{(3)} ,
{\vec p} , {\vec q}  \,
\right) \,
\vert \chi^ m \rangle \, , 
\label{norm2}
\end{eqnarray}
where 
\begin{eqnarray}
\beta_{t T}^{(i)} \left( {\vec p} , {\vec q} \, \right) \;
= \;
\phi_{t T}^{(i)} \left( {\vec p} , {\vec q} \, \right) \;
+ \;
\alpha_{t T}^{(i)} \left( {\vec p} , {\vec q} \, \right) \;
\label{norm3}
\end{eqnarray}
and the $ \alpha_{t T}^{(i)} \left( {\vec p} , {\vec q} \, \right) $ 
functions are given by Eq.~(\ref{alpha}).
Because of the orthogonality of the 
$ \mid \left( t \frac12 \right) T \rangle$ 
and
$ \mid  \vec p\, \vec q \, \rangle $ states,
inserting (\ref{norm2}) into (\ref{norm1}) yields
\begin{flalign}
& 1 = 3 
\sum_{t T} 
\int d^3 p 
\int d^3 q 
\sum_{ i,j = 1}^ 8
\phi_{t T}^{(i)} \left( {\vec p} , {\vec q} \, \right) \;
\beta_{t T}^{(j)} \left( {\vec p} , {\vec q} \, \right) \;
& \nonumber \\
& \langle \chi^ m \mid \,
O_i^\dagger \left(
{\vec \sigma}_{(1)} ,
{\vec \sigma}_{(2)} ,
{\vec \sigma}_{(3)} ,
{\vec p} , {\vec q}  \,
\right) \,
O_j \left(
{\vec \sigma}_{(1)} ,
{\vec \sigma}_{(2)} ,
{\vec \sigma}_{(3)} ,
{\vec p} , {\vec q}  \,
\right) \,
\vert \chi^ m \rangle \, , &
\label{norm5}
\end{flalign}
where the conjugate operators
$O_i^\dagger \left( \vec \sigma_{(1)}, \vec \sigma_{(2)}, \vec \sigma_{(3)}, \vec p, \vec q \, \right) $
are
\begin{eqnarray}
O_1^\dagger & = &  O_1 \, , \nonumber \\
O_2^\dagger & = &  O_2\, ,  \nonumber \\
O_3^\dagger & = &  -O_3 \, ,  \nonumber \\
O_4^\dagger & = &  \frac{1}{ \sqrt{2}} \Big[ -i \vec \sigma( 23)  \cdot \left( {\bf \hat p} \times {\bf \hat q} \, \right)
 - \left( \vec \sigma_{(1)} \times \vec \sigma( 23) \, \right) \cdot ({\bf \hat p} \times {\bf \hat q} \, ) \Big] \, ,  \nonumber \\
O_5^\dagger & = &  -\frac{1}{i} \Big[ \vec \sigma( 23)  \cdot ({\bf \hat p} \times {\bf \hat q} \, ) \ + \ \frac{ i}{2} \left( \vec \sigma_{(1)} \times
\vec \sigma( 23) \right)   \cdot ({\bf \hat p} \times {\bf \hat q} \, ) \, \Big]\, ,   \nonumber \\
O_6^\dagger & = &  O_6\, ,  \nonumber \\
O_7^\dagger & = &  O_7\, ,  \nonumber \\
O_8^\dagger & = &  O_8  \, .
\end{eqnarray}

If we sum both sides of (\ref{norm5}) over $m$, then we are led to 
\begin{eqnarray}
2 = 3 
\sum_{t T} 
\int d^3 p 
\int d^3 q 
\sum_{ i,j = 1}^ 8
\phi_{t T}^{(i)} \left( {\vec p} , {\vec q} \, \right) \;
\beta_{t T}^{(j)} \left( {\vec p} , {\vec q} \, \right) \;
\tilde{C}_{ij}\left( \vec p \vec q, {\vec p}  {\vec q} \, \right) \, ,
\label{norm6}
\end{eqnarray}
where
\begin{eqnarray}
\tilde{C}_{ij}\left( \vec p \vec q, {\vec p}  {\vec q} \, \right) \, = \,
\sum_{m}  \langle \chi^ m|  O_i^\dagger \left( \vec
\sigma_{(1)}, \vec \sigma_{(2)}, \vec \sigma_{(3)}, \vec p, \vec q \, \right) \nonumber \\
O_j \left( \vec \sigma_{(1)}, \vec \sigma_{(2)}, \vec \sigma_{(3)}, {\vec p} , {\vec q} \, \right)
|\chi^ m \rangle \, .
\label{norm7}
\end{eqnarray}
Now all the three functions in (\ref{norm6}) are scalar, so we end up with the following 
normalization condition:
\begin{eqnarray}
1 = 12 \, \pi^2 \,
\sum_{t T} 
\sum_{ i,j = 1}^ 8
\int\limits_0^\infty dp \, p^2 \,
\int\limits_0^\infty dq \, q^2 
\int\limits_{-1}^{1} dx \,
\phi_{t T}^{(i)} \left( p, q , x \right) \;
\beta_{t T}^{(j)} \left( p , q , x \right) \;
\tilde{C}_{ij}\left( x \right) \, ,
\label{norm8}
\end{eqnarray}
where (\ref{pqvec}) has been used.

Using the same $ \tilde{C}_{ij}$ coefficients, we can
calculate the expectation value of the kinetic energy,
$\langle E_{kin} \rangle $. We start from 
\begin{eqnarray}
\langle E_{kin} \rangle = \langle \Psi \mid H_0 \mid \Psi \rangle = 
\langle \psi  ( 1 + P ) \mid H_0 \mid (1 + P ) \psi \rangle = 
3 \, \langle \psi  \mid H_0 \mid  \Psi \rangle \, ,
\label{ekin}
\end{eqnarray}
since $H_0$ and $P$ commute. 
Because the kinetic energy operator is diagonal 
for the basis states 
$ \mid \left( t \frac12 \right) T \rangle$ 
and
$ \mid  \vec p\, \vec q \, \rangle $
and does not depend on any angles, we simply get  
\begin{eqnarray}
\langle E_{kin} \rangle = 12 \, \pi^2 \,
\sum_{t T} 
\sum_{ i,j = 1}^ 8
\int\limits_0^\infty dp \, p^2 \,
\int\limits_0^\infty dq \, q^2 \,
\left( 
\frac{p^2}{m} + \frac{3 q^2}{4 m} 
 \, \right)
\nonumber \\
\times \int\limits_{-1}^{1} dx \,
\phi_{t T}^{(i)} \left( p, q , x \right) \;
\beta_{t T}^{(j)} \left( p , q , x \right) \;
\tilde{C}_{ij}\left( x \right) \, .
\label{ekin2}
\end{eqnarray}
Next we calculate the expectation value of the 2N potential energy, 
$\langle E_{pot}^{2N} \rangle$:
\begin{eqnarray}
\langle E_{pot}^{2N}  \rangle \equiv
\langle \Psi \mid V_{12} + V_{13} + V_{23} \mid \Psi \rangle = 
3 \, \langle \Psi \mid V_{23} \mid \Psi \rangle \equiv
3 \, \langle \Psi \mid V \mid \Psi \rangle \, ,
\label{epot2N}
\end{eqnarray}
where 
\begin{eqnarray}
\langle \Psi \mid V \mid \Psi \rangle & = &
\sum_{t T} 
\int d^3 p 
\int d^3 q 
\sum_{ k = 1}^8 \,
\beta_{t T}^{(k)} \left( {\vec p} , {\vec q} \, \right) \;
\nonumber \\
& \times & \sum_{t' T'} 
\int d^3 p' 
\int d^3 q' 
\sum_{ i = 1}^8 \, 
\beta_{t' T'}^{(i)} \left( {\vec p'} , {\vec q'} \, \right) \;
\nonumber \\
& \times & \langle \chi^ m \mid \,
O_k^\dagger \left(
{\vec \sigma}_{(1)} ,
{\vec \sigma}_{(2)} ,
{\vec \sigma}_{(3)} ,
{\vec p} , {\vec q}  \,
\right) \,
\nonumber \\
& \times & \langle \vec p\, \vec q \, \mid 
\langle \left( t \frac12 \right) T \mid
V
\mid \left( t' \frac12 \right) T' \rangle
\mid  \vec p'\, \vec q' \, \rangle
\nonumber \\
& \times & O_i \left(
{\vec \sigma}_{(1)} ,
{\vec \sigma}_{(2)} ,
{\vec \sigma}_{(3)} ,
{\vec p'} , {\vec q'}  \,
\right) \,
\vert \chi^ m \rangle \, .
\label{epot2N3}
\end{eqnarray}
Using Eqs.~(\ref{vexpand}) and (\ref{isov2})
we get 
\begin{eqnarray}
\langle \vec p\, \vec q \, \mid 
\langle \left( t \frac12 \right) T \mid
V
\mid \left( t' \frac12 \right) T' \rangle
\mid  \vec p'\, \vec q' \, \rangle \, = 
\nonumber \\
\delta \left(  
\vec q - \vec q' \,
\right) \, \delta_{t t'} \, 
\sum_{ j = 1}^6 \,
v_{ t T T'}^ {
(j)} \left( \vec p, \vec p' \, \right) \; 
w_j (\vec \sigma_{(2)}, \vec \sigma_{(3)}, \vec p, \vec p') \, .
\label{epot2N4}
\end{eqnarray}
This leads to 
\begin{eqnarray}
\langle \Psi \mid V \mid \Psi \rangle \, = \,
\sum_{t T}
\int d^3 p
\int d^3 q
\sum_{ k = 1}^8 \,
\beta_{t T}^{(k)} \left( {\vec p} , {\vec q} \, \right) \;
\nonumber \\
\sum_{ T'}
\int d^3 p'
\sum_{ i = 1}^8 \,
\beta_{ t T'}^{(i)} \left( {\vec p'} , {\vec q} \, \right) \;
v_{ t T T'}^ {(j)} \left( \vec p, \vec p' \, \right) \; 
\nonumber \\
\times \langle \chi^ m \mid \,
O_k^\dagger \left(
{\vec \sigma}_{(1)} ,
{\vec \sigma}_{(2)} ,
{\vec \sigma}_{(3)} ,
{\vec p} , {\vec q}  \,
\right) \,
\nonumber \\
\times  w_j (\vec \sigma_{(2)}, \vec \sigma_{(3)}, \vec p, \vec p') \, 
O_i \left(
{\vec \sigma}_{(1)} ,
{\vec \sigma}_{(2)} ,
{\vec \sigma}_{(3)} ,
{\vec p'} , {\vec q}  \,
\right) \,
\vert \chi^ m \rangle \, .
\label{epot2N5}
\end{eqnarray}
Again we sum both sides of Eq.~(\ref{epot2N5}) over $m$ 
and can write $\langle \Psi \mid V \mid \Psi \rangle$ 
in terms of a new set of scalar coefficients, $ \tilde{L}_{kji}$:
\begin{eqnarray}
\langle \Psi \mid V \mid \Psi \rangle \, = \, \frac12 \, 
\sum_{t T}
\int d^3 p
\int d^3 q
\sum_{ k = 1}^8 \,
\beta_{t T}^{(k)} \left( {\vec p} , {\vec q} \, \right) \;
\nonumber \\
\sum_{ t T'}
\int d^3 p'
\sum_{ j = 1}^6 \,
\sum_{ i = 1}^8 \,
\beta_{ T'}^{(i)} \left( {\vec p'} , {\vec q} \, \right) \;
v_{ t T T'}^ {(j)} \left( \vec p, \vec p' \, \right) \;
\tilde{L}_{k j i} (\vec p \vec q,  \vec p \vec p',  \vec p' \vec q \, )
\, ,
\label{epot2N6}
\end{eqnarray}
where
\begin{eqnarray}
\lefteqn{\tilde{L}_{k j i} (\vec p \vec q,  \vec p' 
\vec p'',  \vec p''' \vec q''') = }\cr
& & \sum_{m} \langle \chi^ m|  O^\dagger_k ( \vec \sigma_{(1)}, \vec
\sigma_{(2)}, \vec \sigma_{(3)}, \vec p, \vec q)  w_j(\vec \sigma_{(2)},
\vec \sigma_{(3)}, \vec p', \vec p'' )\cr
 & & \times O_{i} ( \vec \sigma_{(1)}, \vec
\sigma_{(2)}, \vec \sigma_{(3)}, \vec p''', \vec q''') 
| \chi^ m \rangle \, .
\label{ltilde}
\end{eqnarray}
Taking into account the scalar nature of all the functions that appear 
in Eq.~(\ref{epot2N6}), we arrive at the final expression:
\begin{eqnarray}
\langle \Psi \mid V \mid \Psi \rangle \, = \, 4 \, \pi^2 \,
\sum_{t T}
\int\limits_0^\infty dp \, p^2 \,
\int\limits_0^\infty dq \, q^2 \,
\int\limits_{-1}^{1} dx 
\nonumber \\
\times \sum_{ T'}
\int\limits_0^\infty dp' \, {p'}^2 \,
\int\limits_{-1}^{1} dx' \,
\int\limits_{0}^{2 \pi} d \phi' 
\,
\sum_{ k = 1}^8 \,
\sum_{ j = 1}^6 \,
\sum_{ i = 1}^8 \,
\beta_{t T}^{(k)} \left( p, q, x \right) \;
\nonumber \\
\times \beta_{t T'}^{(i)} \left( p', q, x' \right) \;
v_{ t T T'}^ {(j)} \left(  p, p' , y  \right) \;
\tilde{L}_{k j i} ( p , x, p', x', \phi' \, )
\, .
\label{epot2N7}
\end{eqnarray}
Here we use Eq.~(\ref{pqvec}) and assume that
the $ \vec p'$ vector, which appears in Eq.~(\ref{epot2N6}),
is given as 
\begin{eqnarray}
 \vec p' = \left( p' \sqrt{ 1 - {x'}^2 } \cos \phi'  ,  p' \sqrt{ 1 - {x'}^2 } \sin \phi' , p' x' \,
 \right) \, ,
\label{pprimevec}
\end{eqnarray}
which yields 
\begin{flalign}
& x' = {\bf \hat p'} \cdot {\bf \hat q} \, , \nonumber & \\
& y \equiv  {\bf \hat p} \cdot {\bf \hat p' }  = x x' \, + \, \sqrt{ 1 - x^2 } \; \sqrt{ 1 - {x'}^2 } \; \cos \phi' \, . &
\label{y}
\end{flalign}
Note that the 3N operators from Eq.~(\ref{o18}) are
given in terms of unit vectors, so 
no dependence on the magnitude of the $\vec q$ 
vector appears in the $\tilde{L}_{k j i} $ coefficients.

Let us consider also the last part of the 3N Hamiltonian 
and
calculate the expectation value of the 3N potential energy, 
$\langle E_{pot}^{3N} \rangle$:
\begin{eqnarray}
\langle E_{pot}^{3N}  \rangle \equiv
\langle \Psi \mid V^{(1)} + V^{(2)} + V^{(3)} \mid \Psi \rangle = 
3 \, \langle \Psi \mid V^{(1)} \mid \Psi \rangle \, ,
\label{epot3N}
\end{eqnarray}
where 
\begin{eqnarray}
\langle \Psi \mid V^{(1)} \mid \Psi \rangle \, = \,
\sum_{t T} 
\int d^3 p 
\int d^3 q 
\sum_{ k = 1}^8 \,
\beta_{t T}^{(k)} \left( {\vec p} , {\vec q} \, \right) \;
\nonumber \\
\times \sum_{t' T'} 
\int d^3 p' 
\int d^3 q' 
\sum_{ i = 1}^8 \, 
\beta_{t' T'}^{(i)} \left( {\vec p'} , {\vec q'} \, \right) \;
\nonumber \\
\times \langle \chi^ m \mid \,
O_k^\dagger \left(
{\vec \sigma}_{(1)} ,
{\vec \sigma}_{(2)} ,
{\vec \sigma}_{(3)} ,
{\vec p} , {\vec q}  \,
\right) \,
\nonumber \\
\times \langle \vec p\, \vec q \, \mid 
\langle \left( t \frac12 \right) T \mid
V^{(1)}
\mid \left( t' \frac12 \right) T' \rangle
\mid  \vec p'\, \vec q' \, \rangle
\nonumber \\
\times O_i \left(
{\vec \sigma}_{(1)} ,
{\vec \sigma}_{(2)} ,
{\vec \sigma}_{(3)} ,
{\vec p'} , {\vec q'}  \,
\right) \,
\vert \chi^ m \rangle \, .
\label{epot3N2}
\end{eqnarray}
With (\ref{V3N}) 
$ \langle \vec p\, \vec q \, \mid 
\langle \left( t \frac12 \right) T \mid
V^{(1)}
\mid \left( t' \frac12 \right) T' \rangle
\mid  \vec p'\, \vec q' \, \rangle
\, = \, 
\delta_{ T T'} \, 
V^{(1)}_{t t' T}  \left(
\vec p , \vec q  ,  \vec p' ,  \vec q' \,
\right) \, , $
we write Eq.~(\ref{epot3N2}) as
\begin{eqnarray}
\langle \Psi \mid V^{(1)} \mid \Psi \rangle \, = \,
\sum_{t T} 
\int d^3 p 
\int d^3 q 
\sum_{ k = 1}^8 \,
\beta_{t T}^{(k)} \left( {\vec p} , {\vec q} \, \right) \;
\nonumber \\
\times \sum_{t' } 
\int d^3 p' 
\int d^3 q' 
\sum_{ i = 1}^8 \, 
\beta_{t' T}^{(i)} \left( {\vec p'} , {\vec q'} \, \right) \;
\nonumber \\
\times \langle \chi^ m \mid \,
O_k^\dagger \left(
{\vec \sigma}_{(1)} ,
{\vec \sigma}_{(2)} ,
{\vec \sigma}_{(3)} ,
{\vec p} , {\vec q}  \,
\right) \,
\nonumber \\
\times V^{(1)}_{t t' T}  \left(
\vec p , \vec q  ,  \vec p' ,  \vec q' \,
\right) \,
O_i \left(
{\vec \sigma}_{(1)} ,
{\vec \sigma}_{(2)} ,
{\vec \sigma}_{(3)} ,
{\vec p'} , {\vec q'}  \,
\right) \,
\vert \chi^ m \rangle \, .
\label{epot3N3}
\end{eqnarray}
Repeating standard steps (summing both sides of Eq.~(\ref{epot3N3}) over $m$ 
and introducing  yet another set of scalar coefficients, $ \tilde{E}_{k  i}^{t t' T} $)
leads us to:
\begin{eqnarray}
\langle \Psi \mid V^{(1)} \mid \Psi \rangle \, = \, \frac12 \, 
\sum_{t T}
\int d^3 p
\int d^3 q
\sum_{ k = 1}^8 \,
\beta_{t T}^{(k)} \left( {\vec p} , {\vec q} \, \right) \;
\nonumber \\
\times
\sum_{ t'}
\int d^3 p' 
\int d^3 q' 
\sum_{ i = 1}^8 \,
\beta_{t' T}^{(i)} \left( {\vec p'} , {\vec q'} \, \right) \;
\tilde{E}_{k  i}^{t t' T}  (\vec p \vec q,  \vec p \vec q \vec p'  \vec q',  \vec p' \vec q' \, )
\, ,
\label{epot3N6}
\end{eqnarray}
with
\begin{eqnarray}
\lefteqn{\tilde{E}_{k i}^{t t' T}  (\vec p \vec q, \vec p \vec q  \vec p' 
\vec q',  \vec p' \vec q') = }\cr
& & \sum_{m} \langle \chi^ m|  O^\dagger_k ( \vec \sigma_{(1)}, \vec
\sigma_{(2)}, \vec \sigma_{(3)}, \vec p, \vec q)  
\,
V^{(1)}_{t t' T}  \left(
\vec p , \vec q  ,  \vec p' ,  \vec q' \,
\right) \cr
& & \times O_{i} ( \vec \sigma_{(1)}, \vec
\sigma_{(2)}, \vec \sigma_{(3)}, \vec p', \vec q') 
| \chi^ m \rangle \, .
\label{etilde}
\end{eqnarray}
In the last step we use the fact that the functions in Eq.~(\ref{epot3N6}) are scalar
and reduce the number of integrals in that equation:
\begin{flalign}
& \langle \Psi \mid V^{(1)} \mid \Psi \rangle \, = \, 4 \, \pi^2 \,
\sum_{t T}
\int\limits_0^\infty dp \, p^2 \,
\int\limits_0^\infty dq \, q^2 \,
\int\limits_{-1}^{1} dx  &
\nonumber \\
& \times
\sum_{ t'}
\,
\int\limits_0^\infty dp' \, {p'}^2 \,
\int\limits_{0}^{\pi} d\theta_{p'} \sin \theta_{p'}   \,
\int\limits_{0}^{2 \pi} d \phi_{p'} 
\,
\int\limits_0^\infty dq' \, {q'}^2 \,
\int\limits_{0}^{\pi} d\theta_{q'} \sin \theta_{q'}   \,
\int\limits_{0}^{2 \pi} d \phi_{q'} &
\nonumber \\
& \times
\sum_{ k = 1}^8 \,
\sum_{ i = 1}^8 \,
\beta_{t T}^{(k)} \left( p, q, x \right) \;
\beta_{t' T}^{(i)} \left( p', q', z \right) \; & \nonumber \\
& \times
\tilde{E}^{t t' T}_{k  i} ( p , q, x,  p', \theta_{p'}, \phi_{p'},   q', \theta_{q'}, \phi_{q'}  \, )
\, , &
\label{epot3N7}
\end{flalign}
where
\begin{eqnarray}
 z \equiv {\bf \hat p'} \cdot {\bf \hat q'}  = 
 \sin \theta_{p'}  \sin \theta_{q'} \cos \left( \phi_{p'} - \phi_{q'} \, \right) 
+  \cos \theta_{p'}  \cos \theta_{q'}  \, .
\label{z}
\end{eqnarray}
This simply follows from Eq.~(\ref{pqvec}) and from representing 
the $\vec p'$ and $\vec q'$ vectors in a spherical coordinate system as
\begin{eqnarray}
 \vec p' = \left( 
                   p' \sin \theta_{p'} \cos \phi_{p'} \, , \,
                   p' \sin \theta_{p'} \sin \phi_{p'} \, , \,
                   p' \cos \theta_{p'} \, \right) \, , \nonumber \\
 \vec q' = \left( 
                   q' \sin \theta_{q'} \cos \phi_{q'} \, , \,
                   q' \sin \theta_{q'} \sin \phi_{q'} \, , \,
                   q' \cos \theta_{q'} \, \right) \, .
\label{pqprimevec}
\end{eqnarray}

\section{Numerical performance}

It is clear that our method to prepare the full 3N bound state from the Faddeev amplitude,
where we applied the permutation operator "to the right", may also be used to build 
iterations required for Eq.~(\ref{faddeev2}). One first applies the $(1+P)$ operator to the result of the previous
iteration, $| \psi^{(n-1)} \rangle $, and gets an auxiliary state  
$ | \psi_{aux} \rangle  =  (1+P)  | \psi^{(n-1)} \rangle $. 
Then the operators $V$ and $V^{(1)}$ are separately applied onto 
$| \psi_{aux} \rangle $ to yield $ |  \psi_1 \rangle  = V | \psi_{aux} \rangle $ 
and $ | \psi_2 \rangle = V^{(1)} |  \psi_{aux} \rangle$.
The result of the new iteration, $| \psi^{(n)} \rangle $, is just obtained 
as $G_{0} \left(| \psi_1 \rangle + | \psi_2 \rangle \right)$.
This scheme appears most efficient for the Faddeev equation with a 3N force 
and we used it in our calculations. 
We would like to emphasize that the 
$\tilde{C}_{ij}$, 
$\tilde{L}_{kji}$ and
$\tilde{E}_{ki}^{tt'T}$
coefficients introduced in the previous section
can also be used in this new treatment of the Faddeev equation~(\ref{faddeev2}). 
One can equally well project from the left with 
$ \langle \chi^ m|  O_k ( \vec \sigma_{(1)}, \vec \sigma_{(2)}, \vec
\sigma_{(3)}, \vec p, \vec q) $ 
or with
$ \langle \chi^ m|  O^\dagger_k ( \vec \sigma_{(1)}, \vec \sigma_{(2)}, \vec
\sigma_{(3)}, \vec p, \vec q) $. 
In the latter case we arrive at
\begin{eqnarray}
\lefteqn{\sum_{ i = 1}^ 8 \tilde{C}_{ki}(\vec p \vec q,\vec p \vec q )  \;
\phi_{ tT}^ {(i)} ( \vec p,\vec q)  =} \cr
& & \frac{1}{ E-\frac{p^2}{m} - \frac{3}{4m} q^2}
\int d^3 p' \sum_{T'} \sum_{ j = 1}^6 \sum_{ k' = 1}^8
v_{ t T T'}^{ (j)}  (\vec p, \vec p')
 \tilde{L}_{k j k' } ( \vec p \vec q, \vec p \vec p' , \vec p' \vec q )\beta_{ tT'}^ {(k')}
( \vec p',\vec q )\cr
&+& \frac{1}{ E-\frac{p^2}{m} - \frac{3}{4m} q^2}
\int d^3 p' d^3 q' \sum_{t'}
\sum_{ k' = 1}^ 8 \tilde{E}_{kk'}^{ t t' T} ( \vec p \vec
q, \vec p \vec q \vec p' \vec q', \vec p' \vec q')\beta_{ t'T}^ {(k')}
( \vec p',\vec q')\, .
\label{eq:7900}
\end{eqnarray}
The $\beta_{ tT}^ {(k)} ( \vec p ,\vec q ) $
coefficients in Eq.~(\ref{eq:7900}) are the scalar coefficients 
for the 
$  \langle \, \vec p \, \vec q \, | \langle \left( t \frac12 \right) T | (1 + P ) \psi \rangle $
expansion
and depend on $\phi_{ tT}^ {(i)} ( \vec p,\vec q) $ 
as outlined in Eqs.~(\ref{alpha}) and (\ref{norm3}).
The dependence on all relevant variables in Eq.~(\ref{eq:7900}) can be made explicit 
as in the previous section. 

In the case where only 2N forces are included, we obtain results based on three different iteration schemes:
\begin{enumerate}
\item $ | \psi \rangle = G_0 V ( 1 + P ) \, | \psi  \rangle$ with $P$ applied "in both directions" as described in Eq. (\ref{eq:790})
of Sect.~2,
\item $ | \psi \rangle = G_0 t P \, | \psi \rangle $ as described in detail in Ref.~\cite{2N3N},
\item $  |\psi \rangle = G_0 V ( 1 + P ) \, | \psi  \rangle$ as outlined above.
\end{enumerate}
Note that in all the three cases we use the same set of 3N operators~(\ref{o18})
and 2N operators~(\ref{eq:3}). The first set of 3N operators introduced in Ref.~\cite{2N3N} 
combined with the 2N operators from~(\ref{eq:3}) led to complex scalar coefficients, which is an unwanted 
feature in the bound state calculations. The numerical realization 
of all the necessary sets of scalar coefficients that appear
in various versions of the Faddeev equation and which are necessary for the full wave function 
were prepared using software for symbolic algebra and exporting the expressions
into the Fortran form. We used {\em Mathematica}$^{\mbox{\textregistered}}$\cite{math}.

In our calculations we restrict ourselves to a chiral NNLO potential from Ref. \cite{evgeny.report}, with $\Lambda = 550 $ MeV$/$c and $\tilde{\Lambda} = 600$ MeV$/$c.
The operator form of such a potential was briefly described in Appendix~C of Ref.~\cite{2n3d}
In the present paper we use only its neutron-proton version, neglecting the $T=\frac32$
component in the triton.


For calculations with two nucleon forces only we use typically 36 $p$ and $q$ Gaussian points
distributed in the $(0, \bar{p})$ and $(0, \bar{q})$ intervals, respectively.
For the particular potential it is fully sufficient to take 
$\bar{p} \approx $ 5 fm$^{-1}$ and $\bar{q} \approx $ 8 fm$^{-1}$.
We work with 40 $x$ Gaussian points to represent the scalar functions
$\phi_{t T}^{(i)} \left( p, q , x \right) $, 
$\alpha_{t T}^{(i)} \left( p , q , x \right) $
and $\beta_{t T}^{(i)} \left( p , q , x \right) $.
The $\phi^\prime$ integrations are carried out using 60 Gaussian points.
We will demonstrate that our calculations lead to fully convergent results.
As already mentioned, the Faddeev equation is solved by an iterative 
Lanczos algorithm \cite{stadler}. The initial, essentially 
arbitrary, amplitudes 
fulfilled, however, the symmetry properties (\ref{eq:94}).
We found that 20 iterations are fully sufficient. 
A typical job with 4096 processes for 20 iterations requires 
in this case less than 15 minutes
on the IBM BLUE GENE/P computer of the JSC, J\"ulich, Germany. Iterations
are in fact very fast - one can prepare and store the essential parts 
of the iteration kernels and deal for each iteration with only two dimensional integrations.


In the much more time-consuming calculations employing a chiral NNLO 3N force \cite{epel02} where 
we deal with the integral kernel with six fold integrations
we use slightly smaller numbers of $p$, $q$ and $x$ points: 32, 32 and 32, 
respectively.
A corresponding job with 20 iterations requires a few hours on the same number of processes.
In our calculations the four fold angular integrations (see for example Eq.~(\ref{epot3N7})) 
are performed using altogether $\approx 2500$ Gaussian points, 
which might leave room for improvement with respect to the accuracy achieved 
in the case without a 3N force.


In order to save computer resources, having at our disposal the PWD version of the bound state code, for a given 3N Hamiltonian, we first solve the Faddeev equation 
and find the triton binding energy using a partial wave representation.
Then for this energy we run our three dimensional calculations and look for 
the eigenstate corresponding to an eigenvalue $\eta$ closest to $1$. Finally,
for the three dimensional approaches, we construct the scalar coefficients
for the full wave function, normalize them as in (\ref{norm8}) 
and calculate expectation values of the kinetic and potential 
energies as described in Eqs.~(\ref{ekin2}), (\ref{epot2N7}) and (\ref{epot3N7}). 
Results for the calculations with the chosen chiral NNLO 2N potential are given 
in Table~\ref{tab1}. The binding energy for this case, given by the partial wave 
based calculations, is $E_{3N}$=-8.299 MeV.

\begin{table}[hp]\centering
\caption{\label{tab1} The closest to $1$ eigenvalue ($\eta$) of the kernel 
of the Faddeev equation for the given 3N 
energy $E_{3N}$=-8.299 MeV, expectation values of the kinetic 
($\langle E_{kin} \rangle$) and 2N potential ($\langle E_{pot}^{2N} \rangle$) 
energies in the triton as well as their sum  ($\langle E_{kin} \rangle$+$\langle E_{pot}^{2N} \rangle$)
for the guiding partial wave based results and for the methods employing the
three dimensional formulation of the Faddeev equation (see text).
}
\begin{tabular}{ccccc}
\hline
        &  PWD     & Scheme~1  & Scheme~2  & Scheme~3 \\
\hline
$\eta$  &  1.0     &  0.99993  & 0.99983   & 0.99961 \\
$\langle E_{kin} \rangle$      &  31.963  &  31.959   &  31.960   &  31.947 \\
$\langle E_{pot}^{2N} \rangle$    & -40.258  & -40.242   & -40.243   & -40.230 \\
$\langle E_{kin} \rangle$+$\langle E_{pot}^{2N} \rangle$ &  -8.295  &  -8.283   &  -8.283   & -8.283 \\
\hline
\end{tabular}
\end{table}

Results of the calculations including additionally the NNLO chiral 3N force 
are shown in Table~\ref{tab2}. Here the calculations using standard 
partial wave representation yield $E_{3N}$=-8.646 MeV for the binding energy.

\begin{table}[hp]\centering
\caption{\label{tab2} The closest to $1$ eigenvalue ($\eta$) of the kernel
of the Faddeev equation for the given 3N
energy $E_{3N}$=-8.646 MeV,
expectation values of the kinetic 
($\langle E_{kin} \rangle$), 2N potential ($\langle E_{pot}^{2N} \rangle$), 3N potential 
($\langle E_{pot}^{3N} \rangle$)
energies in the triton as well as their sum  ($\langle E_{kin} \rangle$+$\langle E_{pot}^{2N} \rangle$+$\langle E_{pot}^{3N} \rangle$)
for the guiding partial wave based calculations and for the method using the
three dimensional formulation of the Faddeev equation with a 3N force (see text around Eq. (\ref{eq:7900})).
}
\begin{tabular}{ccc}
\hline
        &  PWD     & 3D \\
\hline
$\eta$  &  1.0     &  0.99976  \\
$\langle E_{kin} \rangle$      &  33.448 &  33.412   \\
$\langle E_{pot}^{2N} \rangle$    & -41.329  & -41.273  \\
$\langle E_{pot}^{3N} \rangle$    & -0.765  & -0.770    \\
$\langle E_{kin} \rangle$+$\langle E_{pot}^{2N} \rangle$+$\langle E_{pot}^{3N} \rangle$ &   -8.646  &  -8.631 \\
\hline
\end{tabular}
\end{table}

Even if the new results displayed in Tables~\ref{tab1} and \ref{tab2}
show a good agreement with standard partial wave decomposition calculations,
in the following we show more detailed results. We concentrate first on the case
without a 3N force.
 
\begin{figure}[hp]\centering
\includegraphics[width=0.45\textwidth,angle=0]{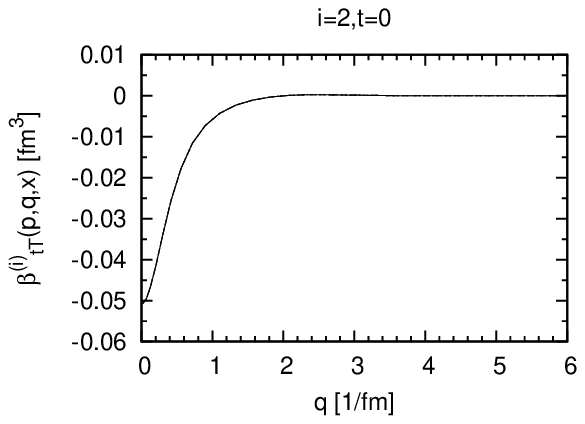}
\includegraphics[width=0.45\textwidth,angle=0]{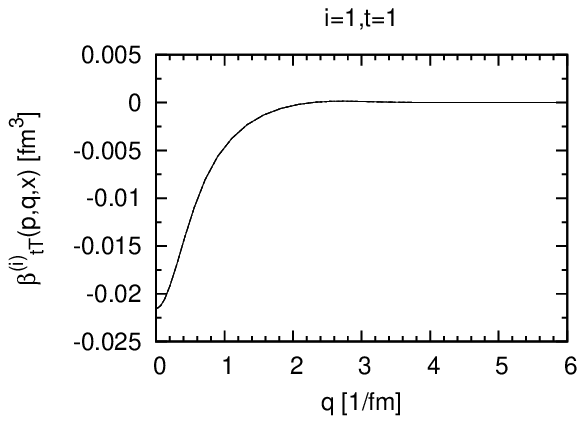}
\includegraphics[width=0.45\textwidth,angle=0]{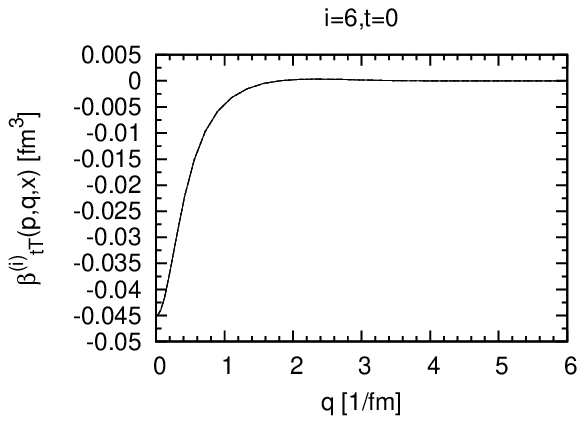}
\includegraphics[width=0.45\textwidth,angle=0]{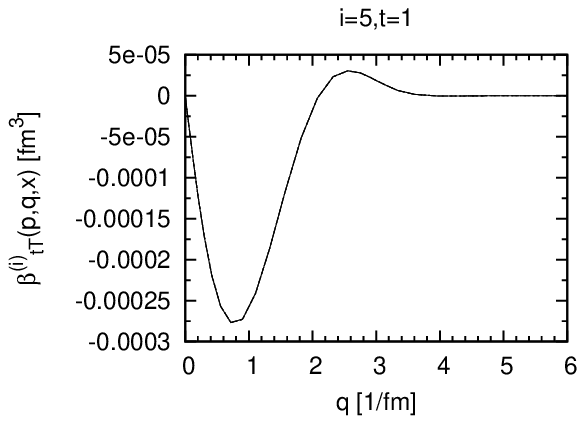}
\includegraphics[width=0.45\textwidth,angle=0]{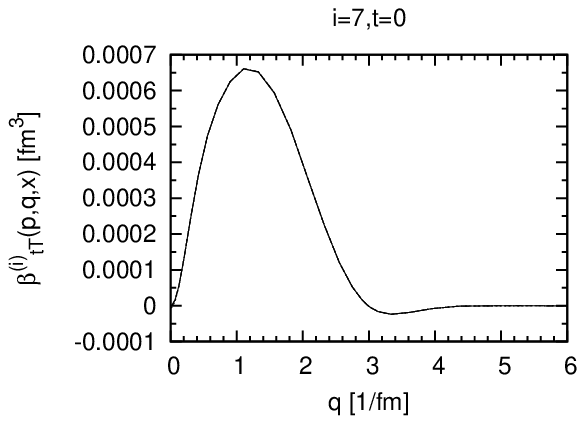}
\includegraphics[width=0.45\textwidth,angle=0]{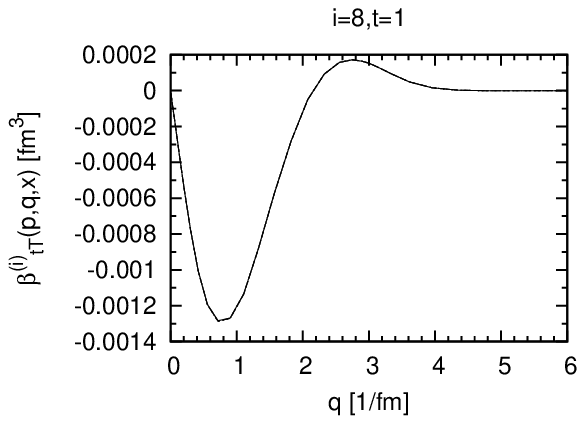}
\caption{Selected scalar coefficients $\beta^{(i)}_{tT} (p,q,x) $ for $T = \frac{1}{2}$ for the full 3N wave function
in the three dimensional form for $p \approx$ 1.4 fm$^{-1}$ and $x \approx$ 0.55 as a function of $q$
obtained with the first (dotted line), second (dashed line) and third (solid line) 
iteration scheme without a 3N force (see text). All three lines essentially overlap. 
}
\label{f3}
\end{figure}
\begin{figure}[hp]\centering
\includegraphics[width=0.45\textwidth,angle=0]{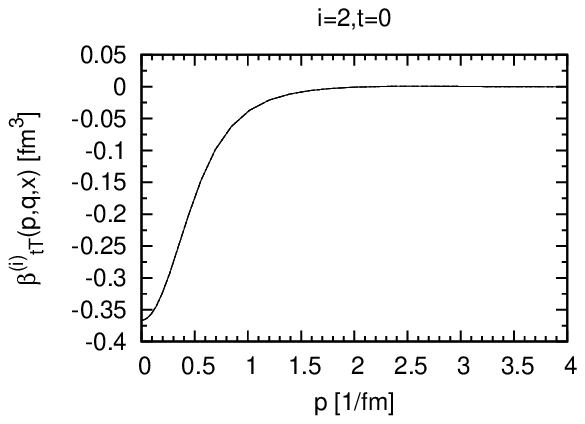}
\includegraphics[width=0.45\textwidth,angle=0]{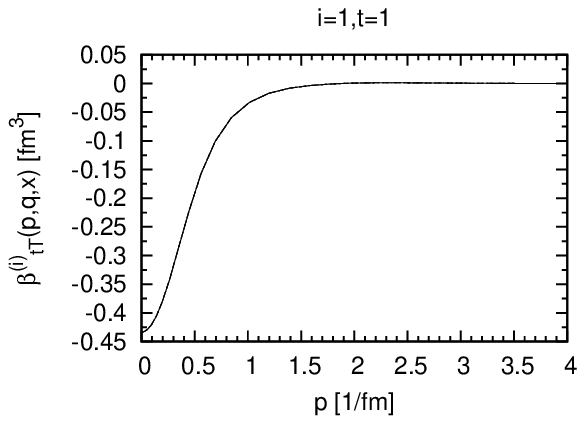}
\includegraphics[width=0.45\textwidth,angle=0]{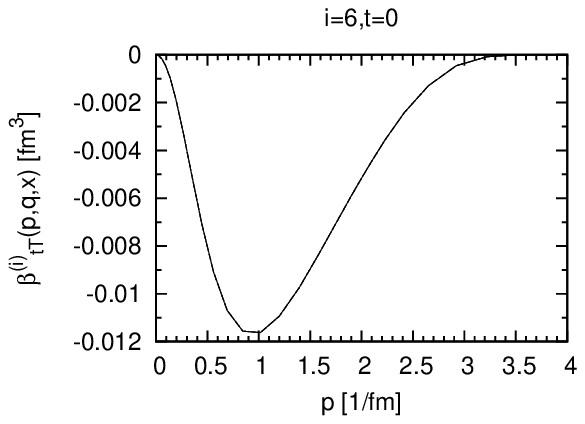}
\includegraphics[width=0.45\textwidth,angle=0]{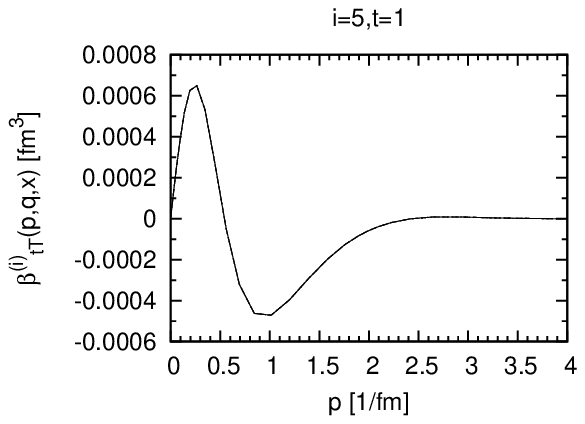}
\includegraphics[width=0.45\textwidth,angle=0]{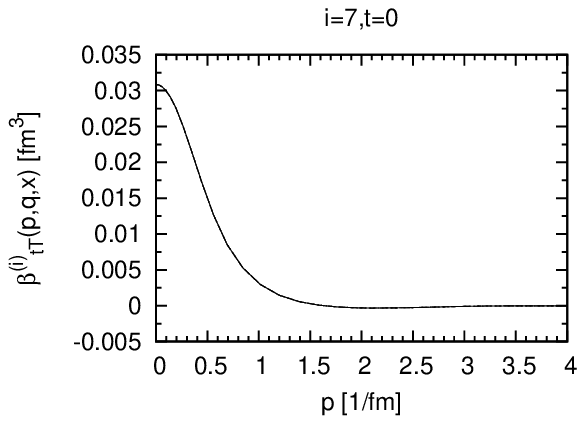}
\includegraphics[width=0.45\textwidth,angle=0]{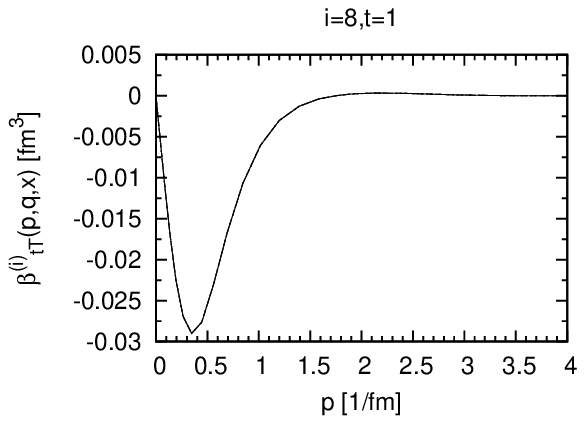}
\caption{Selected scalar coefficients $\beta^{(i)}_{tT} (p,q,x) $ for $T = \frac{1}{2}$ for the full 3N wave function
in the three dimensional form for $q \approx$ 0.72 fm$^{-1}$ and $x \approx$ -0.61 as a function of $p$.
Lines as in Fig.~\ref{f3}.
}
\label{f4}
\end{figure}
\begin{figure}[hp]\centering
\includegraphics[width=0.45\textwidth,angle=0]{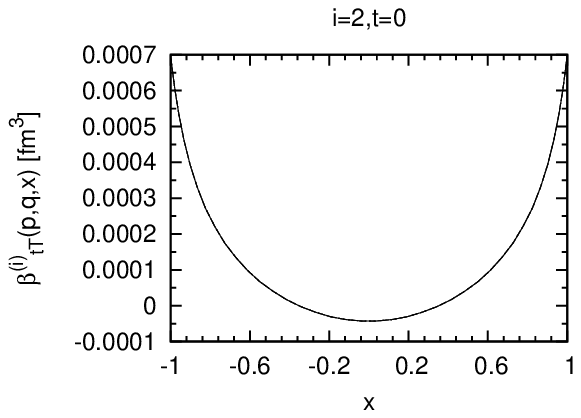}
\includegraphics[width=0.45\textwidth,angle=0]{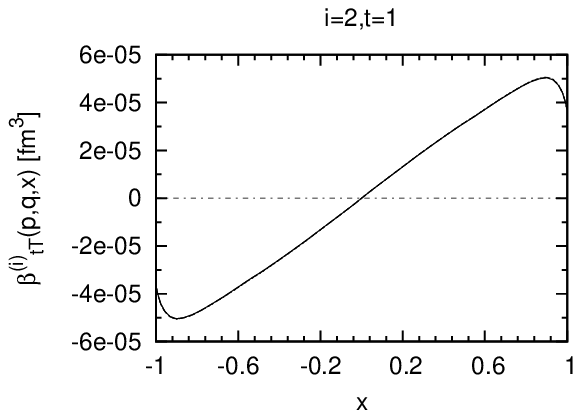}
\includegraphics[width=0.45\textwidth,angle=0]{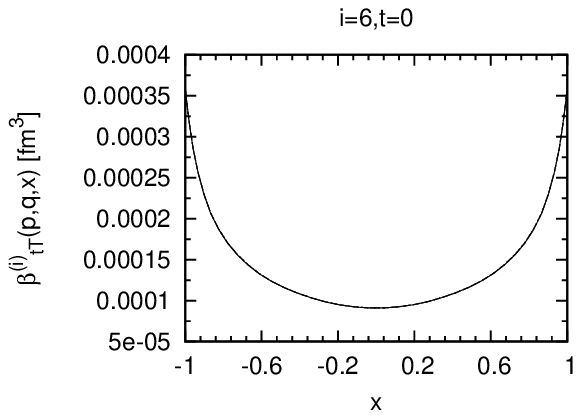}
\includegraphics[width=0.45\textwidth,angle=0]{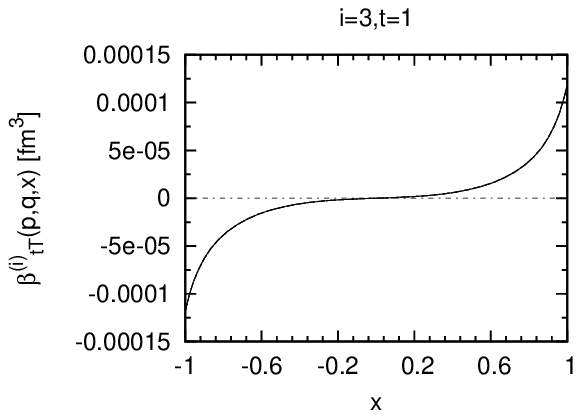}
\includegraphics[width=0.45\textwidth,angle=0]{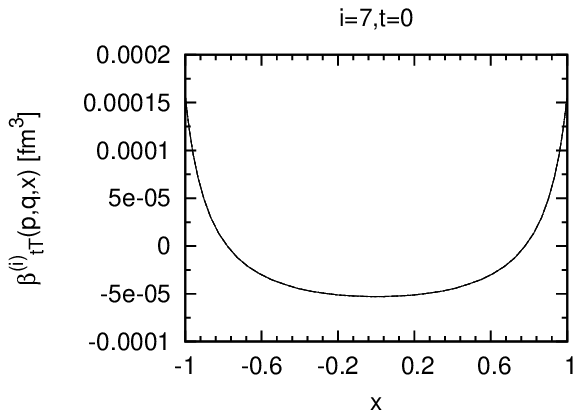}
\includegraphics[width=0.45\textwidth,angle=0]{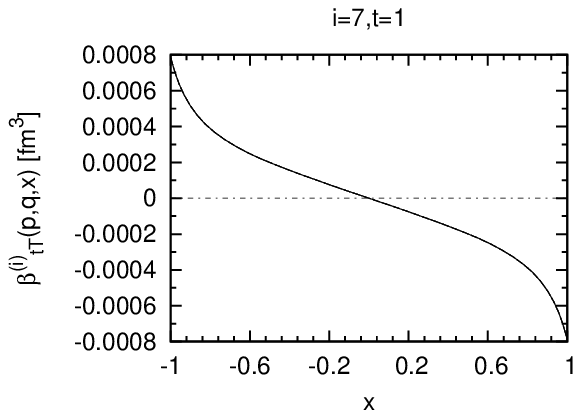}
\caption{Selected scalar coefficients $\beta^{(i)}_{tT} (p,q,x) $ for $T = \frac{1}{2}$ for the full 3N wave function
in the three dimensional form for $p \approx$ 2 fm$^{-1}$ and $q \approx$ 2.6 fm$^{-1}$ as a function of $x$.
Lines as in Fig.~\ref{f3}.
}
\label{f4.5}
\end{figure}

It is interesting to see, if the three quite different 
numerical schemes discussed above lead to the same results. 
They use different sets of scalar coefficients and different 
forms of the permutation operator, $P$. Additionally, the second 
method requires the full off-shell t-matrix, while the other two
methods directly use the 2N potential. Last not least, in the first two methods
the symmetry properties (\ref{eq:94}) were checked after the solution 
had been obtained and in the third case the same symmetry properties 
served to reduce the CPU time. 
In Figs.~\ref{f3}--\ref{f4.5} we show several scalar coefficients for the full
wave function,
$\beta^{(i)}_{tT} (p,q,x)$, first for fixed $p$ and $x$ as a function of $q$, 
then for fixed $q$ and $x$ as a function of $p$ and 
finally for fixed $p$ and $q$ as a function of $x$. In the last case
the symmetry properties (\ref{eq:94}) of the final wave function 
coefficients are clearly visible. In all the shown cases and other (not shown)
the agreement among the three methods is very good.

\begin{figure}[hp]\centering
\includegraphics[width=0.45\textwidth,angle=0]{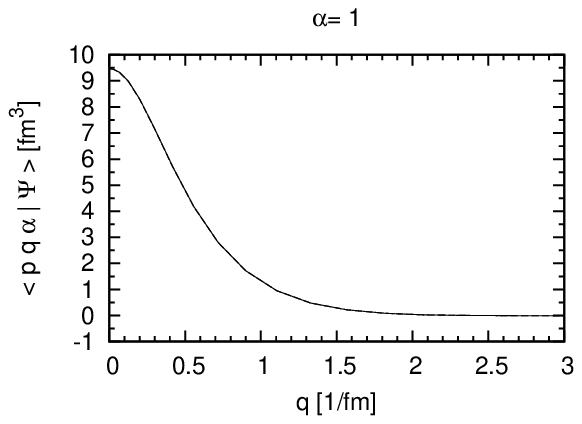}
\includegraphics[width=0.45\textwidth,angle=0]{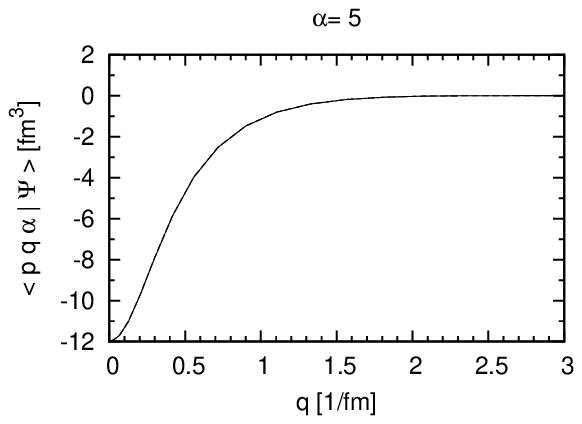}
\includegraphics[width=0.45\textwidth,angle=0]{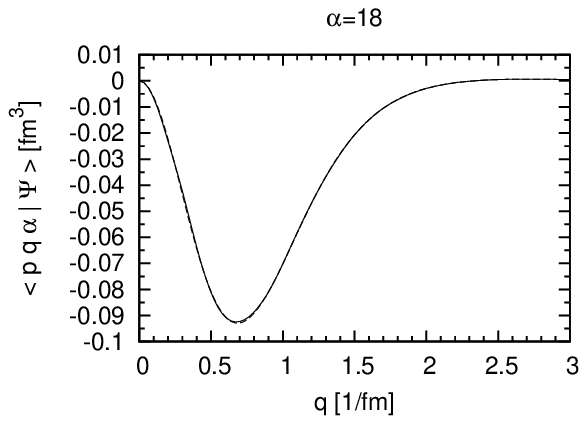}
\includegraphics[width=0.45\textwidth,angle=0]{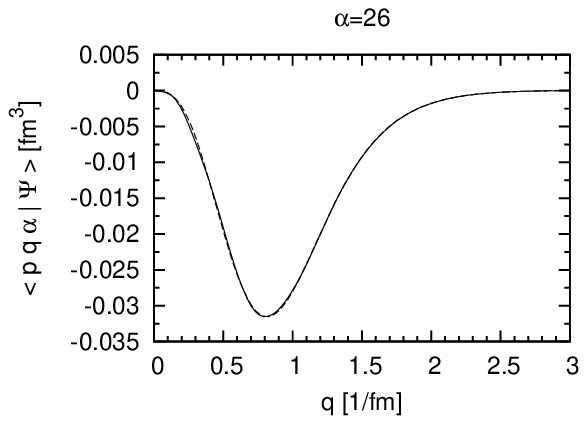}
\caption{The partial wave projected full 3N wave function 
$\langle p q \alpha \frac12 \frac12 ; \left(t \frac12 \right) \frac12 \frac12 | \Psi \frac12 \frac12 \rangle $
for $p \approx$ 0.44 fm$^{-1}$ and four different partial wave states $\alpha$: 
$\alpha=1$ corresponds to $l=0, s=0, j=0, \lambda=0, I=\frac12, t=1$,
$\alpha=5$ gives $l=0, s=1, j=1, \lambda=0, I=\frac12, t=0$,
$\alpha=18$ comprises $l=2, s=1, j=2, \lambda=2, I=\frac52, t=0$
and for $\alpha=26$ we have $l=3, s=1, j=3, \lambda=3, I=\frac72, t=1$
as a function of $q$. The results obtained directly using the partial wave representation 
(dashed line) are compared with the results based on Eq.~(\ref{overlapPWD}). Two lines practically
overlap.
}
\label{f5}
\end{figure}

\begin{figure}[hp]\centering
\includegraphics[width=0.45\textwidth,angle=0]{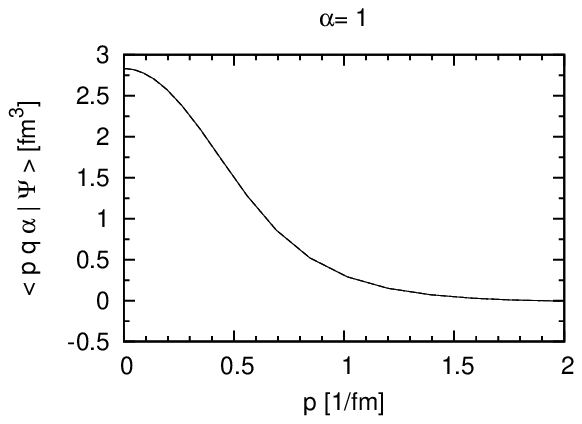}
\includegraphics[width=0.45\textwidth,angle=0]{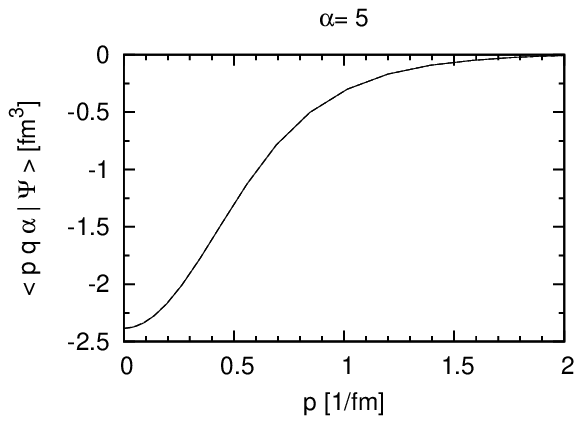}
\includegraphics[width=0.45\textwidth,angle=0]{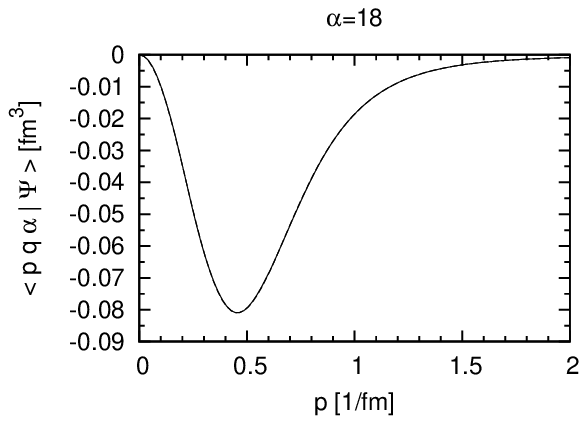}
\includegraphics[width=0.45\textwidth,angle=0]{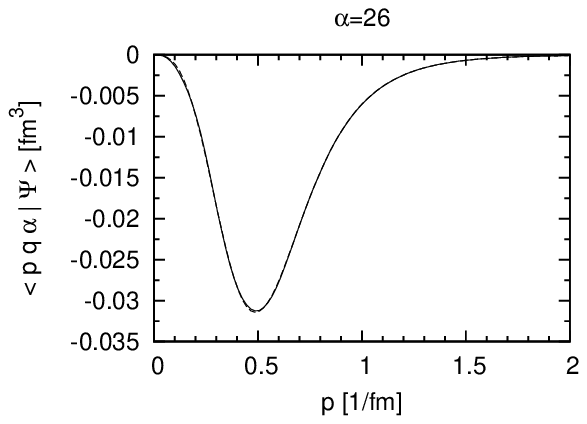}
\caption{The partial wave projected full 3N wave function 
$\langle p q \alpha \frac12 \frac12 ; \left(t \frac12 \right) \frac12 \frac12 | \Psi \frac12 \frac12 \rangle $
for $q \approx$ 0.9 fm$^{-1}$ and the same four different partial wave states $\alpha$
as in Fig.~\ref{f5} as a function of $p$.
}
\label{f6}
\end{figure}

Finally, we show that it is possible to reproduce the standard 
partial wave representation of the full wave function 
starting from the three-dimensional wave function expansion (\ref{norm2}).
To this aim we calculate the overlap \newline
$\langle p q \alpha J M ; \left(t \frac12 \right) T M_T | \Psi m m_t \rangle $,
where $\alpha$ comprises the usual set of discrete quantum numbers for the 3N system,
$\alpha = \{ l, s, j, \lambda, I  \}$ \cite{Gl83}. Here $l$ denotes the orbital 
angular momentum in the two-nucleon subsystem, $s$ is the total spin and $j$ is the
total angular momentum of this subsystem. The orbital angular momentum 
of the third nucleon is coupled with its spin ($\frac12$) to give the total 
angular momentum $I$. Finally, $j$ and $I$ are coupled to yield the total 3N 
angular momentum $J$ with its projection $M$. Using recoupling from the intermediate 
states, where the total 3N angular momentum ($L$) and 
the total 3N spin ($S$) is defined, we arrive at:
\begin{flalign}
& \langle p q \alpha J M ; \left(t \frac12 \right) T M_T | \Psi m m_t \rangle = &\nonumber \\
& \delta_{M_T, m_t} \, \sum\limits_{L,S} \sqrt{ (2 j+1) (2 I +1)  (2 L + 1) (2 S +1 ) } \,
\left\{ 
\begin{array}{ccc}
 l & s & j \\
 \lambda & \frac12 & I \\
  L & S & J 
\end{array}
\right\} & \nonumber \\
& \times \sum\limits_{M_L=-L}^{L} C( L S J ; M_L , M - M_L , M ) \,
\int\limits_0^\pi d\theta_p \sin \theta_p \int\limits_0^{2\pi} d \phi_p \,
\int\limits_0^\pi d\theta_q \sin \theta_q \int\limits_0^{2\pi} d \phi_q & \nonumber \\
& \times 
{\cal Y}^{* \, L M_L}_{ l \lambda} \left( {\hat p} , {\hat q} \, \right) \,
\sum_{ i = 1}^ 8
\beta_{t T}^{(i)} \left( {\vec p} , {\vec q} \, \right) \, &
\nonumber \\
& \times  \langle \left(s \frac12 \right) S M - M_L | 
O_i \left(
{\vec \sigma}_{(1)} ,
{\vec \sigma}_{(2)} ,
{\vec \sigma}_{(3)} ,
{\vec p} , {\vec q}  \,
\right) \,
\vert \chi^ m \rangle \, , &
\label{overlapPWD}
\end{flalign}
where 
\begin{eqnarray}
{\cal Y}^{L M_L}_{ l \lambda} \left( {\hat p} , {\hat q} \, \right) \, \equiv \,
\sum\limits_{m_l=-l}^{l} C( l \lambda L ; m_l , M_L - m_l , M_L ) \,
Y_{ l m_l} \left( {\hat p} \, \right) \, 
Y_{ \lambda M_L - m_l} \left( {\hat q} \, \right) \, ,
\label{overlapPWD2}
\end{eqnarray}
$ C(j_1 j_2 j ; m_1 m_2 m) $ are the Clebsch-Gordan coefficients 
and $ Y_{ l m_l} \left( {\hat p} \, \right) $ denote the spherical harmonics.
The essential parts of the overlap (\ref{overlapPWD}),
\begin{eqnarray}
\langle \left(s \frac12 \right) S M - M_L | 
O_i \left(
{\vec \sigma}_{(1)} ,
{\vec \sigma}_{(2)} ,
{\vec \sigma}_{(3)} ,
{\vec p} , {\vec q}  \,
\right) \,
\vert \chi^ m \rangle \, ,
\label{overlapPWD3}
\end{eqnarray}
can again be prepared with help of the symbolic algebra software. It is also clear that the overlap (\ref{overlapPWD})
can be used for further verification of the three dimensional solution. 
In particular overlaps with
$\mid p q \alpha J M ; \left(t \frac12 \right) T M_T \rangle $
carrying quantum numbers different than the ones of $^3$H should be numerically zero.
We verified very many cases but can show only a few examples 
(see Figs.~\ref{f5}--\ref{f6}), 
which demonstrate that agreement 
with standard partial wave decomposition is perfect not only for the dominant partial waves 
but also for the smaller triton components.
For this comparison we used the three dimensional
wave function obtained with the third method. 

\begin{figure}[hp]\centering
\includegraphics[width=0.45\textwidth,angle=0]{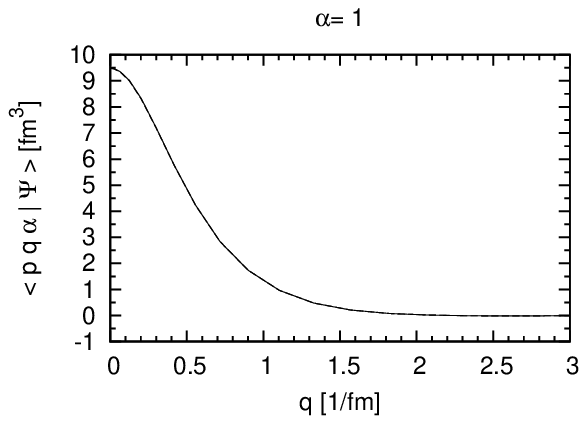}
\includegraphics[width=0.45\textwidth,angle=0]{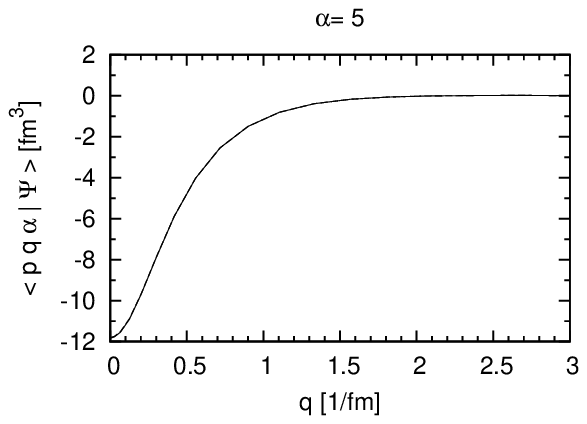}
\includegraphics[width=0.45\textwidth,angle=0]{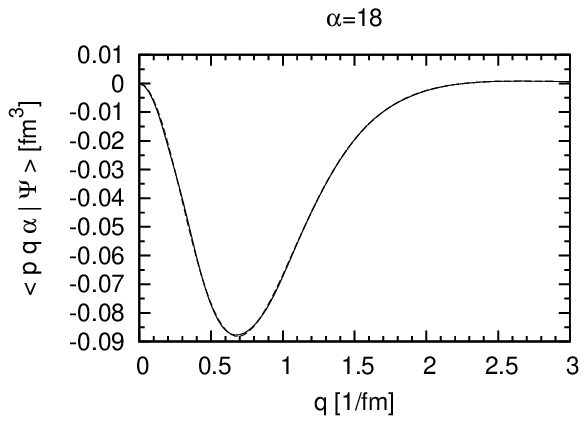}
\includegraphics[width=0.45\textwidth,angle=0]{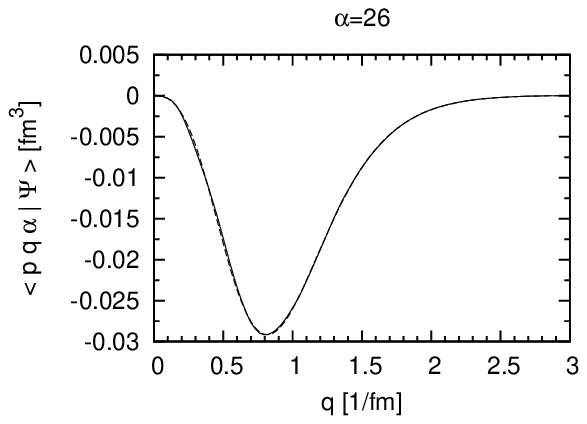}
\caption{The same as in Fig.~\ref{f5} but with inclusion 
of the chiral NNLO 3N force in the both types of calculations. 
}
\label{f7}
\end{figure}

\begin{figure}[hp]\centering
\includegraphics[width=0.45\textwidth,angle=0]{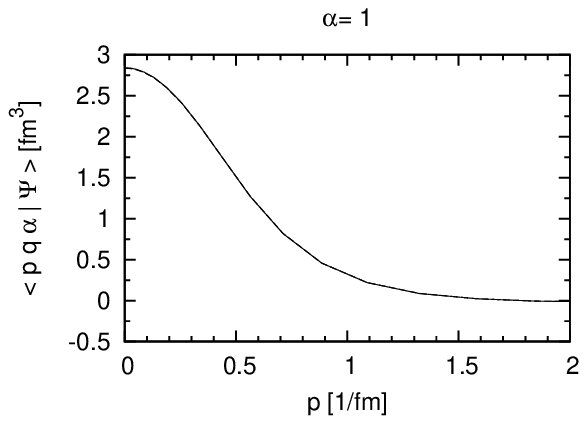}
\includegraphics[width=0.45\textwidth,angle=0]{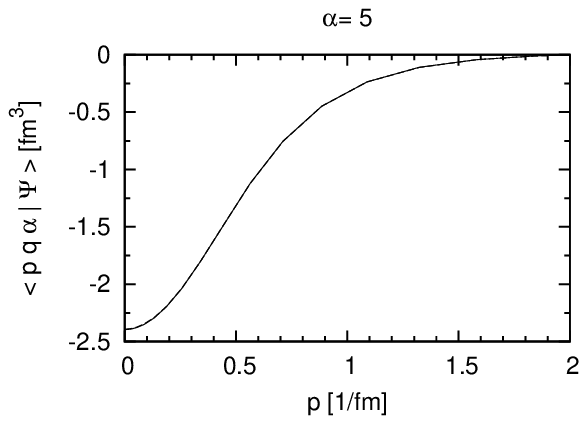}
\includegraphics[width=0.45\textwidth,angle=0]{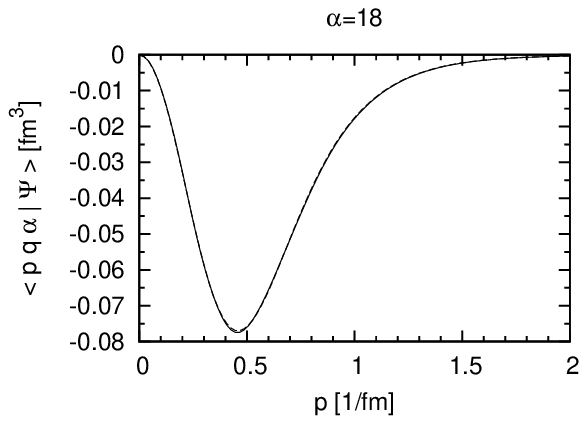}
\includegraphics[width=0.45\textwidth,angle=0]{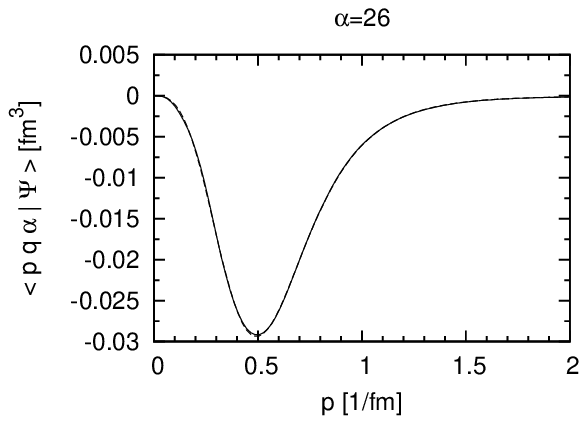}
\caption{The same as in Fig.~\ref{f6} but with inclusion 
of the chiral NNLO 3N force in the both types of calculations. 
}
\label{f8}
\end{figure}

Although the numbers of $p$ and $q$ points in our calculations 
including the chiral NNLO 3N force were smaller than in the case 
where only 2N force were present and the four fold angular integration
in the integral kernel including the 3N force  
might be improved, we show in Figs.~\ref{f7} and~\ref{f8} that 
the projections of the full wave function onto the partial wave
states agree quite well with the standard partial wave representation 
of the 3N bound state. This is remarkable, since the partial wave calculations
were based on Eq.~(\ref{faddeev}) and the three dimensional calculations used 
Eq.~(\ref{faddeev2}).

It is important to notice that the normalization condition 
(analogous to Eq.~(\ref{norm8})) 
\begin{eqnarray}
1 = 4 \, \pi^2 \,
\sum_{t T}
\sum_{ i,j = 1}^ 8
\int\limits_0^\infty dp \, p^2 \,
\int\limits_0^\infty dq \, q^2 
\int\limits_{-1}^{1} dx \,
\beta_{t T}^{(i)} \left( p, q , x \right) \;
\beta_{t T}^{(j)} \left( p , q , x \right) \;
\tilde{C}_{ij}\left( x \right) \, ,
\label{norm88}
\end{eqnarray}
leads to the following normalization 
of the partial wave projections
\begin{eqnarray}
\sum\limits_{\alpha} \,
\int\limits_0^\infty dp \, p^2 \,
\int\limits_0^\infty dq \, q^2
\left( 
\langle p q \alpha \mid \Psi \rangle \,
\right)^2 = 1 \, ,
\label{norm89}
\end{eqnarray}
and provides an additional check of numerics.

\section{Summary and Outlook}

Using an expansion of the Faddeev amplitudes in products of
scalar functions $\phi_{tT}^{(i)}( \vec p, \vec q)$ with
products of scalar expressions in spin and momenta as well as
corresponding expansions of the 2N potential, 2N $t$-operator and 3N forces, 
the Faddeev equations for the 3N bound state can be reformulated into 
a strictly finite set of coupled equations for the amplitudes 
$\phi_{tT}^{(i)}( \vec p, \vec q)$, which depend only on three variables $p$ , $q$ and $\hat{\vec{p}} \cdot \hat{\vec{q}}$.

In this paper we discuss various forms of the Faddeev equation
in the operator form. In comparison to Ref.~\cite{2N3N} 
we propose two new schemes which use directly a 2N potential (without a 3N force)
and a new approach to the Faddeev equation with inclusion 
of a 3N force. For each case we derive corresponding sets of scalar
coefficients. For the three schemes without a 3N force and one scheme 
including a 3N force we provide numerical realizations
based on the chiral NNLO 2N \cite{evgeny.report,2n3d} and 3N \cite{epel02} forces.

The present paper is a continuation of Ref.~\cite{2N3N} and presents 
further elements of the three dimensional formalism. In particular we 
show how to obtain the total 3N bound state from the Faddeev component,
using another form of the permutation operator.
The paper provides also formulas
for the bound state normalization and the expectation values 
of all the parts of the 3N Hamiltonian. 

We show our numerical results for the binding energies 
and display examples of the scalar expansion functions 
for the full 3N bound state. We demonstrate stability and reliability 
of our three dimensional treatment of the bound state Faddeev equation 
by comparing results obtained with the three different methods.
One of the methods required a new fast way to obtain the full 
off-shell $t$-matrix in the operator representation.
We will report on these calculations in a separate paper.
We give all relevant information about the number and distribution of 
the $p$, $q$ and $x$ points used in our calculations.
We are ready for benchmark calculations of the 3N bound state 
in the three dimensional formalism. Our final results for the 
full wave function obtained in the three
dimensional schemes compare very well with the standard results based on the partial
wave decomposition. This is the best proof of the stability and accuracy 
of our numerical performance.


The next step is the extension of our calculations to include the Coulomb force in $^3$He.
The inclusion of this long-ranged force might be easier
in the three dimensional formalism. We also plan to apply 
a three-dimensional operator approach to the problem of
3N scattering and use the formalism outlined in \cite{3NSCATT}.

\section*{Acknowledgments} 
We acknowledge support by the Foundation for Polish Science - MPD program, co-financed by the European Union within the Regional Development Fund. This work was supported by the Polish National Science Center
under Grant No. DEC-2011/01/B/ST2/00578 and partially by the EU
HadronPhysics3 project "Exciting Physics Of Strong Interactions".

One of the authors (JG) would like to thank K. Sagara 
for the hospitality extended to him during his stay
at the Kyushu University and E. Epelbaum 
for the hospitality extended to him during his stay
at the Ruhr-Universit\"at, Bochum.
The numerical calculations
were performed on the supercomputer cluster of the JSC, J\"ulich, Germany.


\bibliographystyle{plain}


\bibliography{thebibliography}

\end{document}